\documentclass[a4paper,11pt]{article}
\pdfoutput=1

\usepackage{jheppub}
\graphicspath{{./figures/}}

\usepackage{xcolor}
\usepackage{xspace}

\newcommand{\BR}{\text{BR}}

 \title{\boldmath Learning Physics at Future $e^-e^+$ Colliders with Machine} 
\author[a]{Lingfeng Li,}
\emailAdd{iaslfli@ust.hk}
\affiliation[a]{Jockey Club Institute for Advanced Study, The Hong Kong University of Science and Technology, Hong Kong S.A.R., P.R.China}
\author[b,c]{Ying-Ying Li,}
\affiliation[b]{Theoretical Physics Department, Fermi National Accelerator Laboratory, Batavia, IL 60510, U.S.A.}
\affiliation[c]{Department of Physics, The Hong Kong University of Science and Technology, Hong Kong S.A.R., P.R.China}
\emailAdd{yingying@fnal.gov}
\author[c]{Tao Liu,}
\emailAdd{taoliu@ust.hk}
\author[c]{Si-Jun Xu}
\emailAdd{sxuaw@connect.ust.hk}
\abstract{Information deformation and loss in jet clustering are one of the major limitations for precisely measuring hadronic events at future $e^-e^+$ colliders. Because of their dominance  in data, the measurements of such events are crucial for advancing the precision frontier of Higgs and electroweak physics in the next decades. We show that this difficulty can be well-addressed by synergizing the event-level information into the data analysis, with the techniques of deep neutral network. In relation to this, we introduce a CMB-like observable scheme, where the event-level kinematics is encoded as Fox-Wolfram (FW) moments at leading order and multi-spectra at higher orders. Then we develop a series of jet-level (w/ and w/o the FW moments) and event-level classifiers, and analyze their sensitivity performance comparatively with two-jet and four-jet events. As an application, we analyze measuring Higgs decay width at $e^-e^+$ colliders with the data of 5ab$^{-1}@$240GeV. The precision obtained is significantly better than the baseline ones presented in documents. We expect this strategy to be applied to many other hadronic-event measurements at future $e^-e^+$ colliders, and to open a new angle for evaluating their physics capability.}

\preprint{FERMILAB-PUB-20-178-T}

\begin{document}
\maketitle
\flushbottom

\section{Introduction}
\label{sec:intro}

The precision frontier of next decades in Higgs and electroweak (EW) physics is expected to be defined by next-generation $e^-e^+$ colliders. The proposed projects so far mainly include~\cite{Taylor:2020} circular machines such as CEPC and FCC-ee, and linear machines such as ILC and CLIC. As a Higgs factory, the CEPC and FCC-ee will operate at $\sqrt{s} = 240$GeV with an integrated luminosity $\sim 5$ab$^{-1}$, in addition to their low-$\sqrt{s}$ runs. About $10^6$ clean Higgs events will be produced during this period together with $\sim 6 \times 10^6~ZZ$, $\sim 8\times 10^7~WW$ and $\sim 3\times 10^8~qq$ events. After this phase, the FCC-ee operation is expected to be upgraded to a mode beyond Higgs factory, with a collection of $\sim 1.7$ab$^{-1}$ data at the $tt$ threshold.  The ILC project aims for $\sim$ 2ab$^{-1}@$250GeV run and $\sim$ 4ab$^{-1}@$500GeV run. As for the CLIC, the planned 380GeV run will collect $\sim 1.5$ab$^{-1}$ data. Then the operation will be upgraded to 1.5 and 3.0 TeV, with a collection of $\sim 2.5$ab$^{-1}$ and $\sim 5$ab$^{-1}$ data, respectively.

\begin{table}[h!]
\centering
\begin{tabular}{ccccc}
\hline 
$N_{q/g}$ & 0 & 2  & 4 & 6 \\ 
\hline 
$e^-e^+ \to WW$ & 11\% & 44\% & 45\% & 0\% \\ 
$e^-e^+ \to ZZ$  & 9\% & 42\% & 48\% & 0\% \\ 
$e^-e^+ \to Zh$ & 3\% & 32\% & 55\% & 11\% \\ 
$e^-e^+ \to \nu\nu h$ & 20\% & 69\% & 11\% & 0\% \\ 
$e^-e^+ \to t\bar t$ & 0\% & 11\% & 44\% & 45\% \\ 
\hline 
\end{tabular} 
\caption{Branching ratios of hadronic modes, for the primary Higgs and electroweak processes in low-$\sqrt{s}$ runs of future $e^-e^+$ colliders. Here $N_{q/g}=0$, $2$, $4$ and $6$ represents the number of (anti-)quarks and gluons in their final states. $\tau$ decays are defined to be leptonic. $W/Z$ decays are assumed to be two-body only.}
\label{tab:jetnumber}
\end{table}

So far, a lot of efforts have been made to explore the prospect of measuring Higgs and EW physics at these $e^-e^+$ colliders~\cite{Fan:2014vta,Banfi:2014sua,dEnterria:2016sca,Fedderke:2015txa,Khanpour:2017cfq,Cai:2016sjz,Chiu:2017yrx,Chen:2018shg,Durieux:2017rsg,Barklow:2017suo,DiVita:2017vrr,Gu:2017ckc,Ge:2016zro,Ge:2016tmm,Ellis:2019zex}. The primary Higgs and electroweak processes for the collider low-$\sqrt{s}$ runs and their branching ratios are summarized in Table~\ref{tab:jetnumber}. Clearly the hadronic modes containing (anti-)quarks or/and gluons are dominant and even overwhelmingly dominant over the purely leptonic ones. Because of this, the baseline sensitivities in documents for many benchmark precision measurements are based on such hadronic modes, with jet-level analysis being generally  applied. One prominent example is the Higgs decay width ($\Gamma_h$) in the Standard Model (SM). At low-$\sqrt{s}$ $e^-e^+$ colliders, its baseline precision is mainly determined by the measurement of $\sigma(\nu\nu h_b)$~\footnote{In this paper we will take a shorthand notation, using the subscript to denote particle decay mode. For example, we will use $h_{q,b,W,Z}$ to denote $h \to qq, bb, WW^*, ZZ^*$, $h_{W_{lq,qq}}$ to denote $h \to WW^* \to l\nu qq, qqqq$ and $h_h$ to denote $h \to qq, bb, gg, \tau\tau$. Additionally, we will use ``$q$'' to denote the quarks of the first two generations and in a general sense, and ``$q_{_3}$'' and ``$q_{_5}$'' to denote $\{u, d, s\}$ and $\{u, d, s, c, b\}$, respectively.} (the Vector-Boson-Fusion (VBF) Higgs production rate with the Higgs decaying into $bb$), with the mainstream method ($i.e.$, Method B defined in Subsec.~\ref{subsec:Hwidth}). At 240GeV, the CEPC and FCC-ee are expected to measure this quantity with a precision of $3.2\%$ and $3.1\%$ (see Table~\ref{tab:obs}), respectively. Combined with the measurements of other intermediate quantities which rely on hadronic data also, this yields a precision of $3.5\%$ for both for the SM $\Gamma_h$ measurement~\cite{An:2018dwb,Abada:2019lih}.      

\begin{figure}[h!]
\centering
\includegraphics[scale=0.16]{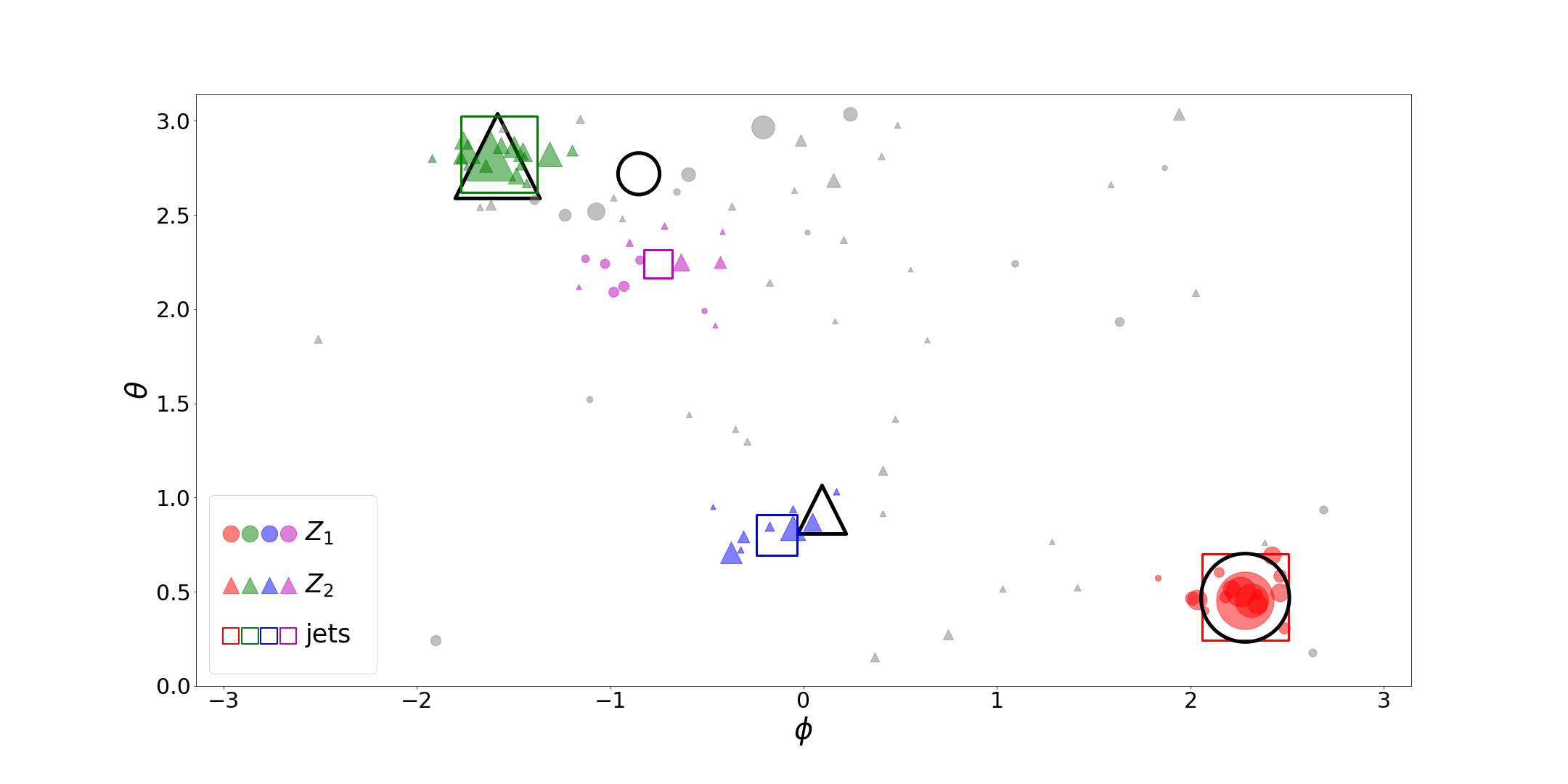}
\includegraphics[scale=0.16]{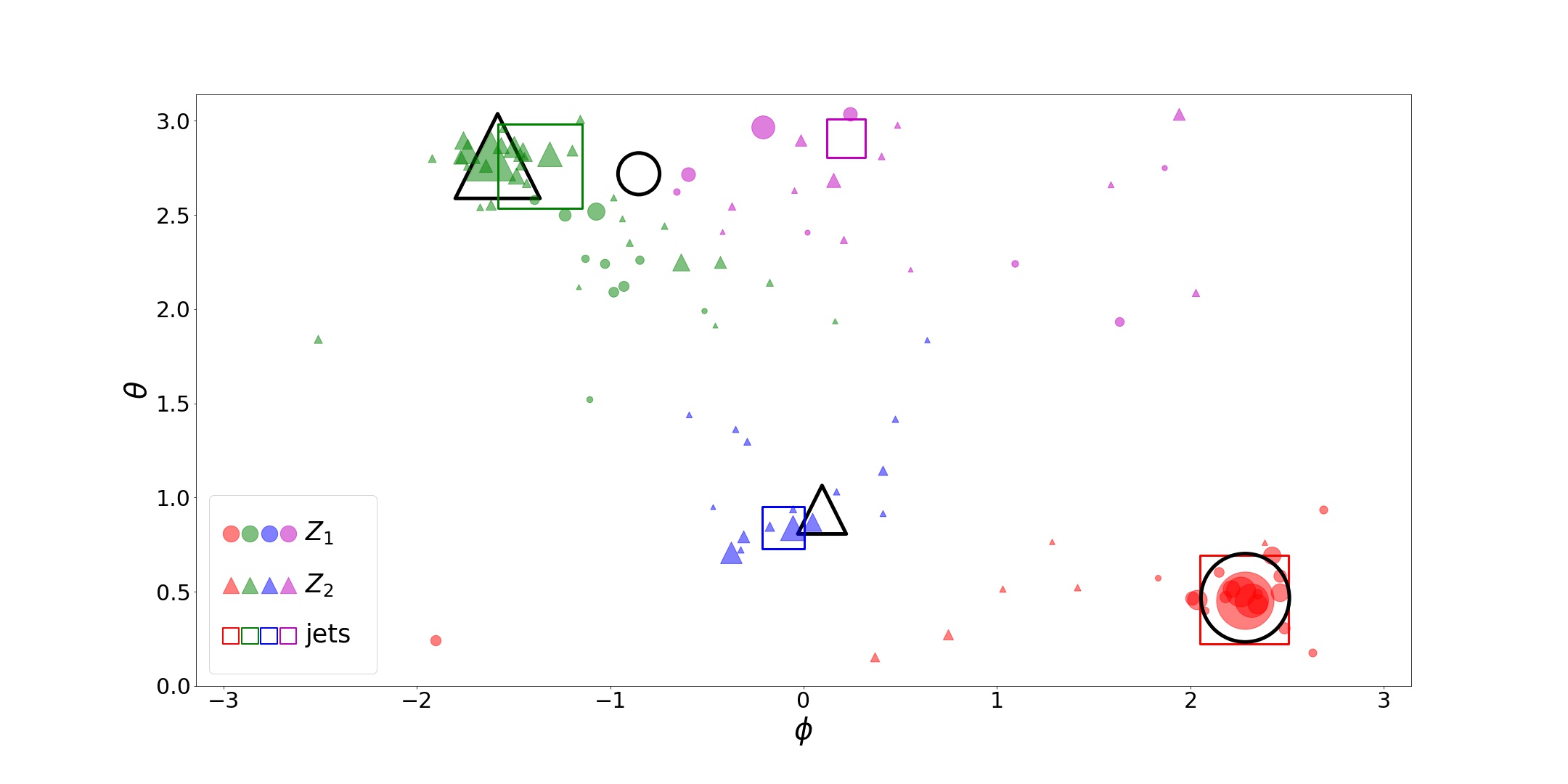}
\caption{An illustration of information deformation and loss in jet clustering. One representative $e^-e^+\to Z_qZ_q$ event is projected to the $\phi-\theta$ plane, without any detector effects. For jet clustering, the anti-$k_t$ algorithm with $\Delta R=0.5$ is applied in the upper panel and the $ee-k_t$ algorithm in the bottom panel. In both panels, each solid symbol represents a particle visible to the detector, with its size scaling with energy, its color denoting the relevant jet, and its shape (circle and triangle) labeling its parent $Z$ boson. The grey symbols in the upper panel represent the particles not clustered to any jets. We use unfilled symbols, $i.e.$, colored boxes and black circles and triangles, to represent the jets and their ancestral quarks, respectively. }
	\label{fig:displayevent1}
\end{figure}

Yet, the precision based on the jet-level analysis is limited for several reasons. First of all, due to the imperfectness of jet clustering algorithms, some visible particles could be clustered into a wrong jet. This becomes especially significant if the jet ancestral partons are collimated, where their hadronizations might badly overlap with each other in space. This effect will deform the jet kinematics from its truth, and may negatively impact the reconstruction of the intermediate particles or events with jets~\footnote{The reconstruction of four momentum of visible particles and their derived quantities such as visible and recoil masses is not influenced by the said information deformation, if no visible particle is missed in jet clustering. But, such high-level observables are often insufficient and even irrelevant for the reconstruction of the whole event. One typical example is the $W_qW_q$ and $Z_qZ_q$ measurement~\cite{Zhu:2018ift}, which will be discussed in Subsec~\ref{subsec:4j}. }. In the performance study of the CEPC detector~\cite{Zhu:2018ift}, this problem is termed as ``jet confusion''. Secondly, the jet clustering in essence is an operation of dimensionality reduction in the feature space of the visible particles. This operation aims reconstructing four momentum of the jet ancestral partons. But, it removes the dimensions reflecting jet substructure and superstructure, generically resulting in a loss of kinematic information. The jet substructure manifests color, electric charge, flavor, etc. of the jet ancestral partons and hence is useful in discriminating, e.g., quark/gluon jets~\cite{Li:2019ufu,Andrews:2019faz,Kasieczka:2018lwf,Komiske:2016rsd,Cheng:2017rdo,Komiske:2018cqr,Andrews:2019faz}.  The jet superstructure is usually formulated if the jet ancestral partons share the same parent particle, where they tend to be showered in a correlated way~\cite{Gallicchio:2010sw}. This structure encodes quantum numbers of these parent particles such as color, spin and CP-property, and is valuable for their collider search. Thirdly, some particles could be either missed by the detector, due to its limited coverage in space, or not clustered to any jets, because of their large distance to the jets. These effects can contribute to the said information deformation and loss also.

To make these problems more explicit,  in Fig.~\ref{fig:displayevent1} we show jet clustering in an $e^-e^+\to Z_qZ_q $ event, using two representative algorithms. The first one is anti-$k_t$ algorithm~\cite{Cacciari:2008gp} which has been extensively applied for data analysis at Large Hadron Collider (LHC). The second one is $ee-k_t$ algorithm~\cite{Catani:1991hj} which has a relatively long history and was originally designed for clean environment at $e^-e^+$ colliders such as the Large Electron-Positron Collider. Unlike the anti-$k_t$, which clusters hard particles first and hence is relatively insensitive to soft radiation and detector noise, the $ee-k_t$ gives a priority to soft particles. This difference allows it to use a priori knowledge on the jet ancestral partons to implement the jet clustering. Explicitly, with the $ee-k_t$ one can request all particles recorded by the detector in each event to be clustered into jets with a given number. For the $e^-e^+\to Z_qZ_q$ event shown in Fig.~\ref{fig:displayevent1}, the anti-$k_t$ clustering is implemented with a jet cone of $\Delta R=0.5$ and the $ee-k_t$ clustering is required to generate four jets. The whole event is projected to the $\phi-\theta$ plane, with no detector simulation being applied. As is shown, the unfilled black circle and triangle on the top of the $\phi-\theta$ plane, which represent jet ancestral partons from different parent $Z$ bosons, are relatively collimated. This results in an overlap between the distributions of solid circles and triangles nearby. Although this unfilled black triangle, together with the unfilled black circle and triangle at the bottom, overlap well with three of the jets at the plane which are denoted by colored boxes, the unfilled black circle on the top is left not close to the fourth jet (magenta box) in both panels. This clearly displays a deformation of kinematic information from the truth for the fourth jet. Indeed, this jet is composed of both solid circles and triangles. Also, as the visible particles are clustered into jets, the information on their correlation and distribution is lost. This information can be partly taken away also by the visible particles which are not clustered to any jets (denoted by grey symbols), if the anti-$k_t$ algorithm is applied. 

The limitations of the jet-level analyses naturally raise the question whether the baseline precisions presented in documents fully reflect the physics potential of future $e^-e^+$ colliders. After all, a significant improvement to many of these baseline precisions would be expected, if the information deformation and loss in jet clustering can be well-addressed. There are two potential methods to solve or partly solve these two problems. The first one is to pursue jet-level analysis by properly incorporating subjet-scale or event-level observables. This method does not solve the problem of information deformation for the jets directly. But it may mitigate its negative impact by incorporating the event-level message. Also, the information lost at jet level can be partly incorporated in this method. The tool of jet-substructure represents such a success which was originally introduced to test QCD~\cite{Ellis:1992qq}. This tool has been extensively applied to searching for boosted heavy objets at LHC. Additionally, a series of event-level observables have been introduced for data analysis at colliders since decades before. These observables fall into two classes roughly. The first class manifest event shape. One prominent example is thrust  
\begin{equation}
T \equiv \max_{\bf{n}_T} \frac{\sum_{i\in {\rm event}} |{\bf p}_i \cdot {\bf n}_T| }{\sum_{i\in {\rm event}} |{\bf p}_i|}
\end{equation} 
introduced in 1970s~\cite{Farhi:1977sg}. Here $\bf{n}_T$ is a unit spacelike vector and defines the thrust axis, and $i$ runs over all visible particles in the event. The thrust was subsequently generalized to many other possibilities~\cite{Stewart:2010tn,PARISI197865,Berger:2003iw,Mateu:2012nk,Catani:1992jc} (for a review, see, e.g.,~\cite{Dasgupta:2003iq,Banfi:2010xy}). It is interesting that most of these event-level observables were originally proposed for the $e^-e^+$ and $e^-h$ events~\cite{Dasgupta:2003iq} rather than the $hh$ ones~\cite{Banfi:2010xy}. Recently, an observable to measure event isotropy was also proposed~\cite{Cesarotti:2020hwb}. Another class encode the event-level information at different angular scales. The most famous example is probably Fox-Wolfram (FW) moments~\cite{Fox:1978vu}. The FW moments were introduced in 1970s also, for analyzing the $e^-e^+$ collision events. They are defined as 
\begin{equation}
H_{AB;l}= \sum_{m=-l}^l H_{AB; l,m} = \frac{4\pi}{2l +1}  \sum\limits_{i,j}   \frac{A_iB_j}{s}  \sum_{m=-l}^l \left (Y_l^m (\Omega_i)^*  Y_l^m (\Omega_j) \right) = \sum\limits_{i,j}\frac{A_iB_j}{s}P_l(\cos \Omega_{ij}) \ .  
\end{equation}
Here $Y_l^m (\Omega_i)$ is spherical harmonics of degree $l$ and order $m$,  $P_l(\cos \Omega_{ij})$ is Legendre polynomials, 
\begin{eqnarray}
\cos \Omega_{ij} = \cos \theta_i \cos \theta_j +  \sin \theta_i \sin \theta_j \cos (\phi_i -\phi_j)
\end{eqnarray}
is the cosine of the included angle between the $i^{\rm th}$ and $j^{\rm th}$ visible particles, and $A$ and $B$ are infrared-safe kinematic variables such as $p_T$, $E$, etc. In this summation $i$ and $j$ run over all visible particles in the event~\footnote{To avoid being distracted from the QCD information, one can modify the definition of the FW moments by excluding the isolated leptons or photons from this summation, as we will do in the analysis of measuring $\sigma(Z_\nu h_{W_{lq}})$ in Subsec.~\ref{ssec:ww}.}. These two classes of observables are both physically intuitive, but less organized or incomplete in representing the event-level kinematics. Another one is to pursue the analysis in a brute-force way, using the event-level data as input. With this method, the problem of information deformation at jet level becomes irrelevant, while the kinematic information at event level could be exploited to the greatest extent for data analysis. Despite this, both methods are confronted with a challenge, $i.e.$, how to efficiently synergize the event-level information into the data analysis, given the complexity of its structure.

The machine learning (ML) techniques based on deep neural network (DNN) bring a great opportunity to address this challenge, due to their revolutionary capability to mine data. This tool became popularized in last two decades for hardware development and big data availability. This motivates us to pursue the study below. Our primary goal is to 
\begin{itemize}
\item provide an angle to evaluate the physics capability of future $e^-e^+$ colliders, which is different from the ones taken in most relevant literatures and documents, by properly synergizing the event-level information into the DNN-based data analysis.  
\end{itemize}
We will develop a set of DNN-based binary classifiers using both methods and apply them to a series of benchmark studies comparatively. Yet, implementing the first method in an organized manner  requires the event-level kinematics to be encoded as a complete or approximately complete set of prioritized observables. So, we would also  
\begin{itemize} 
\item construct an observable scheme to systematically represent the event-level information in each event.
\end{itemize}
By incorporating the observables in such a scheme order by order, we would expect the performance of the jet-level classifiers to approach that of the event-level ones gradually. The  information lost at jet level then could be reconstructed based on these observables. Beyond that, such an information-representing scheme is valuable for seeing into the event-level kinematics and revealing the underlying physics, a task generically difficult for the second method.   

The $e^-e^+$ colliders stand on a better position, compared to hadron colliders, in this regard. They are characterized with negligible pileups, colorless beam, and especially isotropy of event four momentum ($i.e.$, $\vec p_{\rm event} \equiv 0$). This is reminiscent of all-sky Cosmic Microwave Background (CMB) map, and motivates us to introduce a CMB-like observable scheme. In this scheme, the event-level kinematics is encoded as the FW moments at leading order and multi-spectra at higher orders. We will be less ambitious in this paper, and test only to what extent the FW moments of energy can compensate for the information lost at jet level and reduce the performance gap between the jet-level and event-level classifiers. 

This paper is organized as follows. In Sec.~\ref{sec:method} we introduce the CMB-like observable scheme and the strategies for the DNN-based analyses. Then we develop a series of jet-level (w/ and w/o the FW moments) and event-level classifiers, and analyze their performance comparatively with two-jet and four-jet events in Sec.~\ref{sec:BS}. Similar  strategies are subsequently applied to the analysis of measuring $\Gamma_h$ at 240GeV in Sec.~\ref{sec:Hmeasurement}. We summarize our results and take an outlook in Sec.~\ref{sec:summary}.

\section{General Strategies}
\label{sec:method}

\subsection{CMB-like Observable Scheme}
 \label{ssec:CMB}

\begin{figure}[th!]
\centering
\includegraphics[scale=.35]{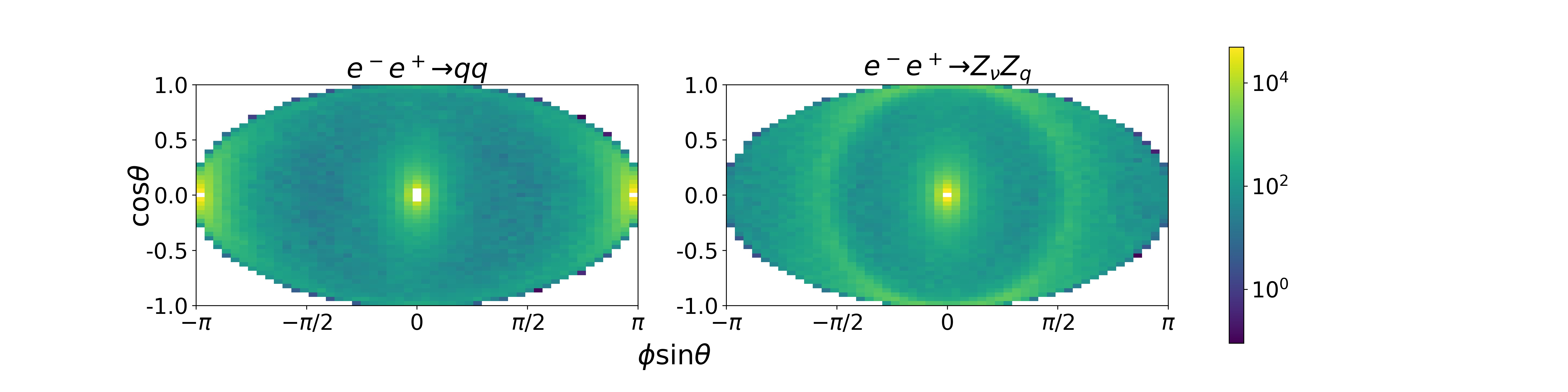}
\caption{Cumulative Mollweide projections of 10000 events: $qq$ (left) and $Z_\nu Z_q$ (right), with the brightness of each cell scaling with the total energy (GeV) of the particle hits received.}
\label{fig:eventimg}
\end{figure}

In this subsection, we will introduce a CMB-like observable scheme to encode the kinematic information in each event. Let us consider first the cumulative Mollweide projection of two classes of hadron-level events: $e^-e^+ \to qq$ and $e^-e^+ \to Z_\nu Z_q$. Here only the visible particles are relevant. We make the projection in the following way: (1) define a Cartesian coordinate system at the collider, with its $z$-axis being along the beam line; (2) rotate the momentum of the most energetic visible particle to be along $x$-axis which points from the paper to outside; and (3) scale the brightness of each cell at the projection sphere with the total energy of the particle hits received. The projections are shown in Fig.~\ref{fig:eventimg}. Both of them demonstrate anisotropic features. In the $e^-e^+ \to qq$ projection, there are two bright points centered at $\{ \phi \sin\theta = 0, \cos\theta =0  \}$ and $\{ \phi \sin\theta = \pm \pi, \cos\theta =0  \}$, respectively. This reflects that before hadronization the two ancestral quarks move oppositely due to momentum conservation. Differently, a pupil-like structure is formed in the $e^-e^+ \to Z_\nu Z_q$ projection, with its circumference and radius being determined by the relative position of these two ancestral quarks and their included angle, respectively. By energy and momentum conservation, the position contour of the second ancestral quark at the $\phi\sin\theta - \cos\theta$ plane is given by 
\begin{eqnarray}
\frac{1}{2} - \frac{1}{2} \sqrt{1-\cos^2\theta} \cos \left( \frac{\phi \sin\theta}{ \sqrt{1-\cos^2\theta}}  \right ) = \frac{m_Z^2}{E_Z^2}.   
\label{eq:inangle}
\end{eqnarray}
This equation predicts the said included angle at $\cos \theta = 0$, $i.e.$, its characteristic value, to be $\sim \phi \sin\theta \sim \pi/2$. With hadronization, the position points and contours for these ancestral quarks are smeared into halos, with their energy density varying spatially. 

\begin{table} [th!]
\centering
\begin{tabular}{c|c}
\hline 
Mollweide projection at $e^-e^+$ colliders & All-sky CMB map \\ 
\hline 
\hline
Projection sphere & Celestial sphere \\
\hline
Equatorial plane & Galactic plane \\
\hline
Energy ($p_T$, timing, charge, $d_0$, etc.) projection & Temperature (polarization) map \\
\hline
Event-level kinematics & Anisotropy \\
\hline
Fox-Wolfram moments & Power spectrum ($TT$, $TB$, $BB$, etc.) \\
\hline
Multi-spectra & Bispectrum, trispectrum, etc. \\
\hline
... ... & ... ... \\
\hline
\end{tabular} 
\caption{Dictionary between the Mollweide projection at $e^-e^+$ colliders and the all-sky CMB map. }
\label{tab:dic}
\end{table}

These observations are reminiscent of the all-sky CMB map where the message on the early Universe is encoded as its power spectrum and multi-spectra. Quite generally, we can build up a dictionary between the Mollweide projection of each $e^-e^+$ collision event and the all-sky CMB map, as is summarized in Table~\ref{tab:dic}. Here the projection sphere plays a role of the celestial sphere in the all-sky CMB map, with its equatorial plane mimicking the disc of Milky Way. The spherical projection of collider observables including energy and momentum, timing, tracker parameters such as charge, impact parameter $d_0$, etc., can be mapped to the all-sky map of the CMB temperature and polarization. Naturally, the event-level kinematics is manifested as the anisotropy of the projection. The relevant information thus can be encoded as the FW moments at leading order and multi-spectra at higher orders, an analogue to the CMB power spectrum and its bispectrum, trispectrum, etc.~\footnote{To ensure their infrared- and collinear-safety, for some observables in this CMB-like scheme such as the charge FW moments and multi-spectra, one needs to properly weight the contribution of each particle using, e.g., its energy or momentum (as was done for defining jet charge in~\cite{Krohn:2012fg}).} Despite these correspondences, it is noteworthy that there exist some important differences between the Mollweide projection at $e^-e^+$ colliders and the all-sky CMB map. First, to high accuracy, the CMB temperature fluctuation is a Gaussian random field, with its non-Gaussian effects being expected to be tiny. In contrast, the Mollweide projection of each event at $e^-e^+$ colliders is physically non-Gaussian. Its multi-spectra thus may contain significant information beyond the FW moments on the event-level kinematics~\footnote{Recall, a Gaussian random field is completely determined by its two-point correlator. All of its higher-order correlators can be expressed in term of the two-point correlator according to Wick's theorem.}. Second, the CMB power spectrum is measured in the universe where we live and hence one realization of all the possible CMBs is recorded only. This limits its measurement precision at large angular scale, causing the notorious ``cosmic variance'' problem. Differently, the collider data sample is typically of large size. The variance of their mean over samples are suppressed for the FW moments at all multipoles, according to the central limit theorem.

\subsection{Machine Learning with Event-level Kinematics}
 \label{ssec:eventmlintro}

In this paper we will apply to our study the two methods to address the information deformation and loss in jet clustering, in a comparative way. Explicitly, we will develop five types of DNN-based binary classifiers in each analysis: 
\begin{itemize}

\item J1 classifier: jet-level, without FW moments and track information; 

\item J2 classifier: jet-level, with $H_{EE;l \leq 50}$ (FW moments of energy with $l\leq 50$) and without track information; 

\item J3 classifier: jet-level, with $H_{EE;l \leq 50}$ and track information; 

\item E1 classifier: event-level, without track information; 

\item E2 classifier: event-level, with track information. 

\end{itemize} 
Among these, J1 will serve as a reference classifier. J2, J3 and E1, E2 classifiers are based on the first and second methods, respectively. We will test the effectiveness of E1 classifier by comparing its performance with J1's. By expectation, E1 classifier should perform better than J1. J2 classifier will tell us to what extent $H_{EE;l \leq 50}$, as part of the leading-order observables in the CMB-like observable scheme, can compensate for the information lost at jet level~\footnote{The information carried by the jets and the FW moments could overlap to some extent. If all observables in this CMB-like scheme are incorporated, we would expect that the jet information become irrelevant.}, and reduce the performance gap between J1 and E1 classifiers. The track observables of secondary vertex (SV) will be incorporated at last in J3 and E2 classifiers. 
 
The event-level classifiers are somewhat related to the end-to-end ones proposed in~\cite{Andrews:2018gew,Andrews:2018nwy}. Yet, instead of using the raw detector response as input for improving particle reconstruction~\cite{Andrews:2018gew,Andrews:2018nwy}, we are more dedicated to addressing the information deformation and loss in jet clustering. Hence we will use the reconstructed particles as input for the analyses. Several difficulties arise in this setup. If the feature space is defined with the momenta of the visible particles in each event, its  dimension is not fixed, due to the fluctuation of the particle number. Also, the dimension of the feature space is generically high for the hadronic events and hard to sort. These complexities could be addressed in several ways with the ML techniques. The first one is to image the events and then apply the ML techniques of image recognition, such as Convolutional Neural Network (CNN)~\cite{Komiske:2019jim,Li:2019ufu,Monk:2018zsb}, for their classification. In this case, the pixel intensity in each image represents the total contribution of the visible  particles hitting this pixel to some kinematic variables such as $E$. The dimension of the input parameters is thus determined not by the particle number, but by the pixel number. Similar techniques of image recognition have been applied to tagging light jets~\cite{Komiske:2016rsd}, boosted $W$ boson~\cite{deOliveira:2015xxd} and top quark~\cite{Kasieczka:2017nvn}, selecting events~\cite{Lin:2018cin,Li:2019ufu,Andrews:2018nwy,Kim:2018cxf,Kim:2019wns}, mitigating pileups at the LHC~\cite{Komiske:2017ubm}, etc. For its directviewing and effectiveness, we will take this method below. The second method is based on Recurrent Neural Network~\cite{Andreassen:2018apy} or its variants such as Recursive Neural Network (RecNN)~\cite{Cheng:2017rdo}. These ML models take inputs from each particle sequentially, yielding a hidden state with fixed dimension. Hence they can deal well with the particle number fluctuation in the events. The third method takes the event-level information as a graph where the graph nodes and edges represent some kind of property of particles and their correlation with each other (e.g, $\Delta R_{ij}$ and $\theta_{ij}$). The hidden state of each node gets updated based on its own properties and the properties of its adjacent edges/nodes. The applications of the graph-based models including the Graph Neural Network and its many variants can be found in~\cite{Ren:2019xhp,Martinez:2018fwc,Farrell:2018cjr,Abdughani:2018wrw,Qu:2019gqs} (for a review, see~\cite{DBLP:journals/corr/abs-1812-08434}). 

Explicitly, we implement all DNNs used in our study in PyTorch~\cite{paszke2017automatic}. We first define three modules of fully-connected neural network (FCN), using jets, FW moments and track observables as their inputs, respectively. Each of them is comprised of 5 hidden layers, with [16, 128, 128, 128, 16] neurons and activation function of ReLU. These modules are then properly connected to construct J1, J2 and J3 classifiers. For the event-level classifiers, image recognition is based on ResNet-50 CNN~\cite{DBLP:journals/corr/HeZRS15}. E1 classifier first passes the event images to the convolution part of a  ResNet-50 network, and then flattens the convolution output to be the input layer of its FCN part. E2 classifier is defined as a FCN with the output neurons of E1 and the track module being its input. The CNN input is taken from a $50\times50$ evenly gridded $\theta-\phi$ plane where the energy intensity is defined at each pixel. The assumed image resolution is consistent with the multipole range of $l\le 50$ for the FW moments incorporated in J2 and J3 classifiers. As a comparison, the proposed CEPC detector template has a granularity ($\phi-\eta$) of $300\times 360$ $(150\times 180)$ in the central region of ECAL (HCAL)~\cite{Chen:2017yel}, and the IDEA detector design of FCC-ee has a dual readout, with the granularity ($\phi-\eta$) of the ECAL/HCAL being $240\times 300$~\cite{IDEACard}. Both of them are finer than the image pixel assumed above. This leaves a space for the simulation setups to absorb the uncertainties arising from the detector granularity which could be achieved. The ResNet-50 network is trained for 50 epochs with a batch size of 512 and a learning rate of 0.0001, using the loss function of binary cross entropy. Adam optimizer~\cite{article} is used for gradient descending of the loss function. All FCNs are trained for 300 epochs with a batch size of 512 and learning rate of 0.001. 

The size of the samples for training and testing each classifier is set to $10^5 + 10^5$ and $5\times 10^4$, respectively, in Sec.~\ref{sec:BS}. In Sec.~\ref{sec:Hmeasurement}, we set it to $3\times 10^5 + 3\times 10^5$ and the real event number for 5ab$^{-1}$ data. For the backgrounds, the training samples are defined based on their real budget after preselection. These samples are simulated with Madgraph5~\cite{Alwall:2011uj} and parton shower with Pythia8~\cite{Sjostrand:2007gs}, unless otherwise specified. For the jet-level classifiers, the visible particles in each event are clustered into jets with $ee-k_T$ algorithm~\cite{Catani:1991hj}, using FastJet~\cite{Cacciari:2011ma}. We assume the detector to be perfect in Sec.~\ref{sec:BS} and use the built-in CEPC-detector~\cite{Ruan:2018yrh} and FCC-ee-IDEA templates~\cite{IDEACard} in DELPHES3~\cite{deFavereau:2013fsa} for the $\Gamma_h$ analysis in Sec.~\ref{sec:Hmeasurement}.  In the $\Gamma_h$ analysis, we also simulate the detector effects on the track observables by smearing the displacement of the SV from the primary vertex (PV) ($d_{\rm vertex}$) by $5\mu m$, $i.e.$, the typical $d_0$ resolution for such detectors~\cite{CEPCStudyGroup:2018ghi,Abada:2019zxq}.

\section{Benchmark Study} 
\label{sec:BS} 

In this section, we will analyze two benchmark scenarios, with each of their events containing two (Subsec.~\ref{subsec:2j}) and four (Subsec.~\ref{subsec:4j}) jets, respectively. 

\subsection{Two-Jet Events}  
\label{subsec:2j}  

In the two-jet benchmark study, we will develop binary classifiers to distinguish between the four classes of $Zh$ Higgs events at $\sqrt{s} =240$ GeV, including: 
\begin{itemize} 

\item $e^-e^+ \to Z_\nu h_b \to  \nu \nu bb$; 

\item $e^-e^+ \to Z_\nu h_g \to \nu \nu gg$;   

\item $e^-e^+ \to Z_\nu h_{q_{_3}} \to \nu \nu q_{_3} q_{_3}$; 

\item $e^-e^+ \to Z_\nu h_{W_{qq}} \to  \nu \nu W_qW_q^* \to \nu\nu qqqq$.

\end{itemize} 
These four classes of events share the same production mechanism, but are differentiated by the number of jet ancestral partons and their color, electric charge, flavor, etc. At (two) jet level, these events benefit very little from the reconstructed jet kinematics (e.g., four momentum) except $b$ tagging for their mutual distinguishing. A large portion of the information on jet ancestral partons, manifested by their showing, gets lost because of jet clustering. J2, J3, E1, E2 classifiers are expected to be able to utilize such lost information to various extents.

\begin{figure}[th!]
\centering
\includegraphics[scale=.3]{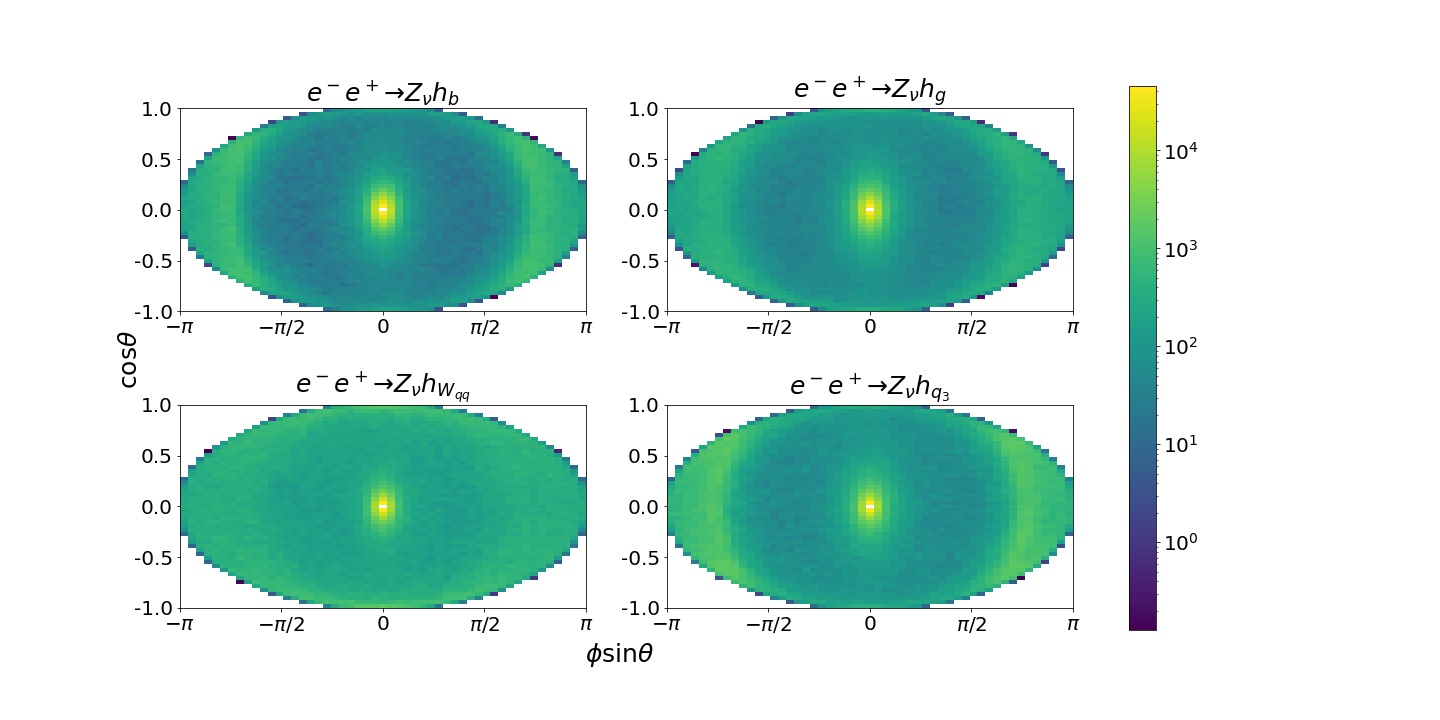}
\caption{Cumulative Mollweide projections of 10000 events: $Z_\nu h_b$ (upper-left), $Z_\nu h_g$ (upper-right), $Z_\nu h_{W_{qq}}$ (bottom-left) and $Z_\nu h_{q_{_3}}$ (bottom-right), with the brightness of each cell scaling with the total energy (GeV) of the particle hits received. }
\label{fig:2jeventimg} 
\end{figure}
We present the cumulative Mollweide projections of these four classes of $Zh$ events in Fig.~\ref{fig:2jeventimg}. In the $Z_\nu h_{g,b,q_{_3}}$ projections, we see a pupil-like structure again. But, compared to that of the $Z_\nu Z_q$ projection in Fig.~\ref{fig:eventimg}, the size of these pupils appears bigger. This is because the parent particle of the two jet ancestral partons for the $Zh$ events (Higgs boson) is heavier than that of the $Z_\nu Z_q$ events ($Z$ boson), which makes its two descendant partons to be less collimated. Analytically, the included angle between these two jet ancestral partons is determined by the formula in Eq.(\ref{eq:inangle}), but with the factor $\frac{m_Z^2}{E_Z^2}$ being replaced with $\frac{m_h^2}{E_h^2}$. The pupil-like structure becomes vague in the $Z_\nu h_{W_{qq}}$ projection. In this case, there exist multiple ways to define the said included angle, upon the jet ancestral-parton pair to consider, and especially, the one contributing the most to the projection predicts a broad distribution for its included angle (to be discussed below). These effects significantly smear such a structure.

\begin{figure}[h!]
	\centering
	\includegraphics[scale=0.295]{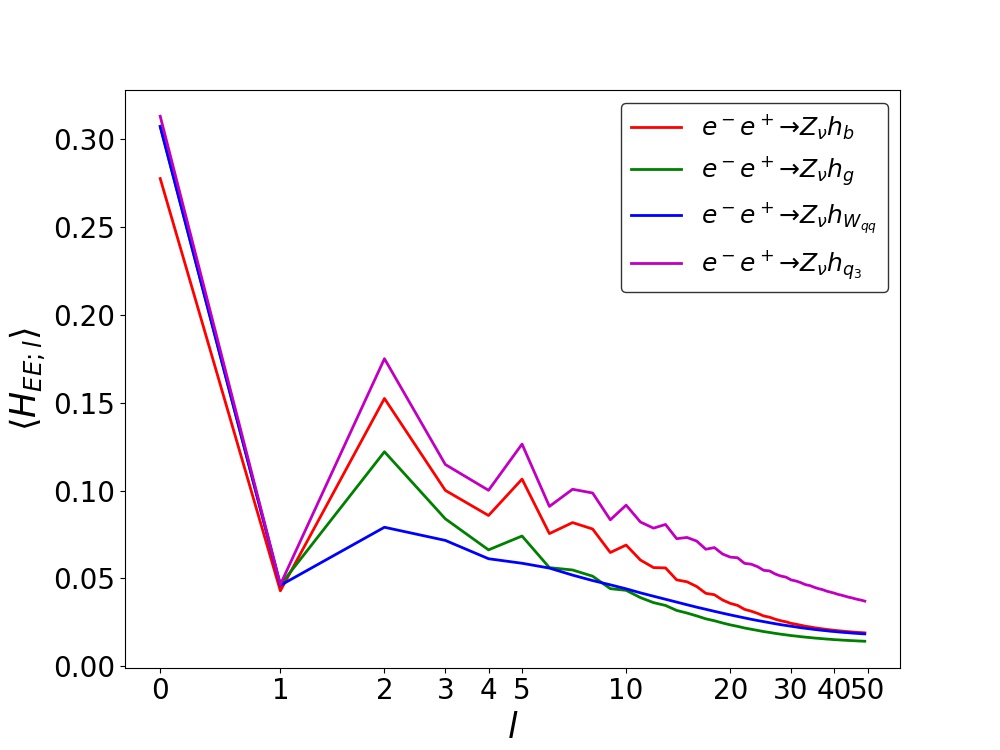}
	\includegraphics[scale=0.295]{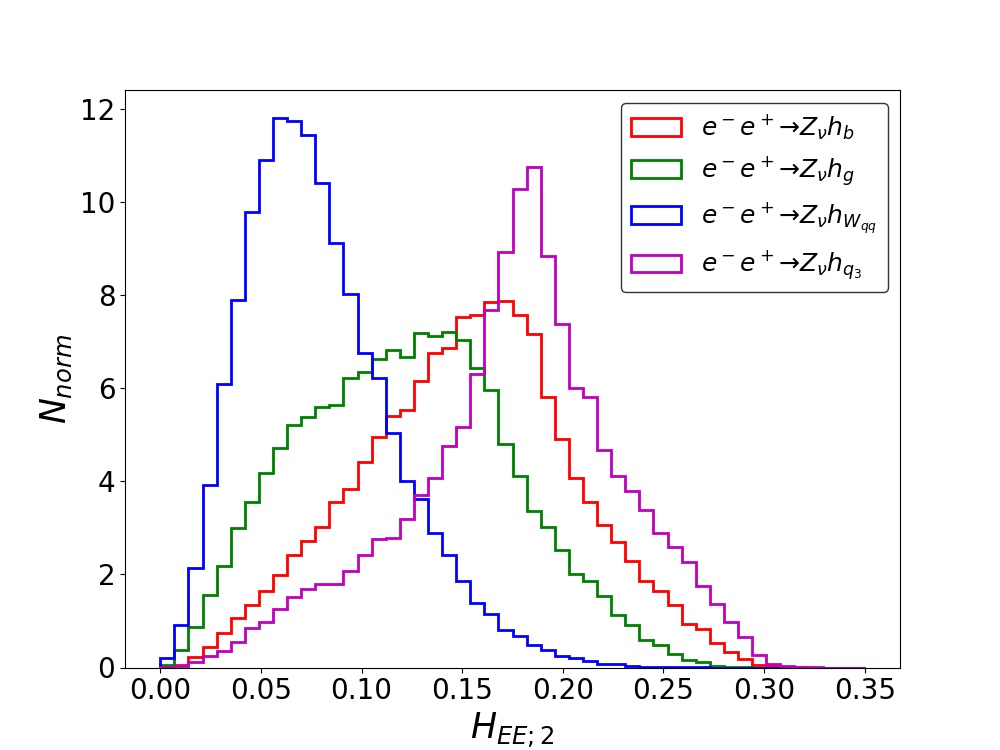}
	\caption{FW spectra of $\langle H_{EE;l} \rangle$ (left) and event distributions of $H_{EE;2}$ (right) for the two-jet samples.}
	\label{fig:FW2j}
\end{figure}
\begin{figure}[h!]
	\centering
	\includegraphics[scale=0.31]{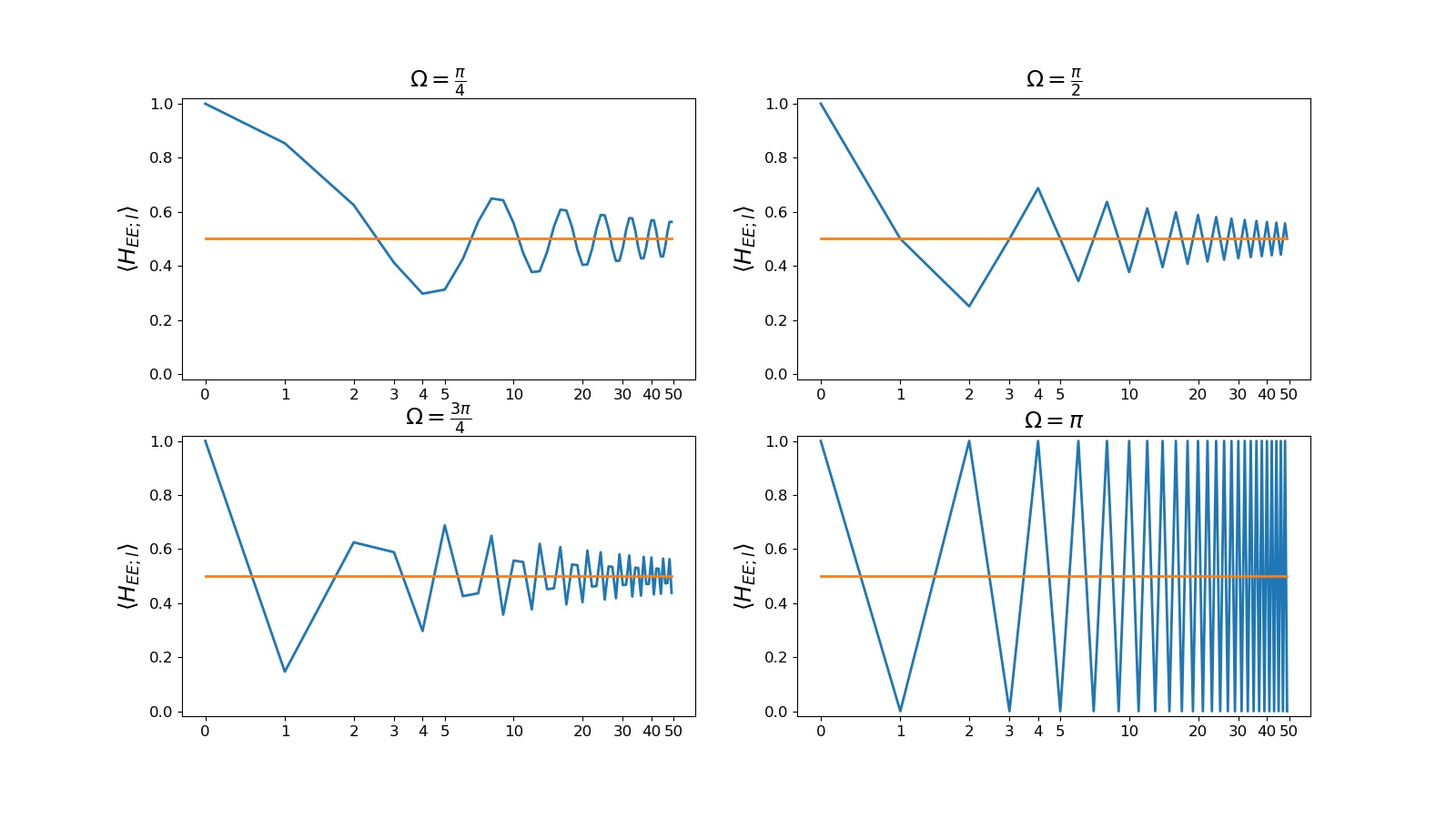}
	\caption{FW spectra of $\langle H_{EE;l}  \rangle$ for a system of two partons with equal energy and varied included angle $\Omega$. The orange lines at $\langle H_{EE;l} \rangle=0.5$ represent the contribution from particle self-correlation to each FW moment.}
	\label{fig:FW1}
\end{figure}
As a manifestation of the event-level kinematics, the FW spectra of $\langle H_{EE;l} \rangle$ and the event distributions of $H_{EE;2}$ for the two-jet samples are presented in Fig.~\ref{fig:FW2j}. Here and below, $\langle H_{EE;l} \rangle$ represents an average of $H_{EE;l}$ over each sample. The $\langle H_{EE;l} \rangle$ spectra are reminiscent of the CMB power spectrum. Similar to the latter, they are characterized by a series of ``acoustic peaks and valleys'' containing rich physical information. To demonstrate this, we plot in Fig.~\ref{fig:FW1} the FW spectra for a system of two jet ancestral partons with equal energy and varied included angle. These spectra are expected to encode the four momenta of these two partons or approximately the jets that they initiate. In comparison, the FW spectra in Fig.~\ref{fig:FW2j} encode not only the four momenta of the jet ancestral partons but also their quantum numbers and even their parent particles'. The physical effects resulting from these quantum numbers (e.g., QCD radiation due to color charge) will deform and smear  the FW spectra of the jet ancestral partons. Such deformation and smearing represent the information lost at jet level generally.  Below are a series of comments on these FW spectra. 
\begin{itemize}

\item The FW moments with odd $l$ are zero for parity-even events, because of $P_l (-x) = (-1)^l P_l (x)$, as happens to the two-parton system with $\Omega = \pi$ in Fig.~\ref{fig:FW1}. In this system, the two-parton correlations contribute to each FW moment 
\begin{eqnarray}
\frac{1}{2}\Big(1+ P_l(\cos\Omega=-1)\Big)=\frac{1}{2}\Big(1+ (-1)^l\Big) \ , \label{eq:pi}
\end{eqnarray} 
yielding a zigzag oscillation in the spectrum along the orange line.  

\item For $0< \Omega < \pi$, the FW oscillation becomes less periodic w.r.t. the multipole $l$. Its amplitude gradually decreases as $l$ increases, due to a suppression caused by Legendre polynomials at high $l$. As $\Omega$ decreases, the enhanced collinearity between the two partons gradually raises the FW spectrum at large angular scales (except $H_{EE;0}$), pushing its first acoustic peak to the high-$l$ end. The FW spectrum becomes a straight line with $\langle H_{EE;l} \rangle \equiv 1$ in the limit of $\Omega = 0$. These effects yield various oscillation patterns for the FW spectrum. Interestingly, the FW moments at low $l$ only are able to determine the nature of $\Omega$ qualitatively. For example, the $\langle H_{EE;1 \le l \le 4} \rangle$ moments for the $Z_\nu h_{g,b,q_{_3}}$ events in Fig.~\ref{fig:FW2j} define a peak at $l=2$ in their respective spectra. This matches approximately with the pattern of the two-parton system with $\Omega = \frac{3}{4}\pi$ in Fig.~\ref{fig:FW1}, and also consists with the indication of the cumulative Mollweide projections in Fig.~\ref{fig:2jeventimg} on the included angle between the two jet ancestral partons in these events. This method will be often used for the discussions below.

\item QCD radiation and hadronization will deform and smear the parton-level FW spectrum. To understand this better, one can split the FW moments into self- and inter-correlation parts, $i.e.$,
\begin{eqnarray}
H_{EE;l} = H_{EE;l}^{\rm self} + H_{EE;l}^{\rm inter}   \ .
\end{eqnarray}
The two terms in Eq.~(\ref{eq:pi}) represent such a splitting also, but at parton level instead. The self-correlation of the visible particles makes a universal contribution 
\begin{eqnarray}
H_{EE;l}^{\rm self} =  \sum_i \frac{ E_i^2}{s}
\end{eqnarray} 
to all-$l$ FW moments in each event. Its magnitude is an anti-measure of the democracy of allocating visible energy among these particles, and is irrelevant to their spatial distribution inside the detector. With more particles (e.g., because of stronger QCD radiation) and fairer energy allocation, this contribution will be reduced.  In Fig.~\ref{fig:FW1}, the orange lines represent such a contribution from two jet ancestral partons with the same energy. If the parton showing is turned on, these orange lines will be shifted downward. This effect results in damped tails for the FW spectra in Fig.~\ref{fig:FW2j}, and ensures the FW moments to be infrared- and collinear-safe theoretically. 

\item The inter-correlation of the visible particles makes an $l$-dependent contribution 
\begin{eqnarray}
H_{EE;l}^{\rm inter}  = \sum\limits_{i,j}^{i\neq j} \frac{ E_i E_j }{s} P_l (\cos \Omega_{ij}) 
\end{eqnarray} 
to the FW moments in each event. $H_{EE;l}^{\rm inter}$ is sensitive to the spatial distribution of these particles inside the detector (except at $l=0$ since $P_0 (\cos \Omega_{ij})  \equiv 1$) and determines the oscillation pattern of the FW spectrum. If these particles are highly collimated, the inter-correlation between any two of them tends to be positive, due to $P_l (\cos (\Omega_{ij} \to 0)) \to 1$. The FW moments at large angular scales will gain more from this since $P_l (\cos \Omega_{ij})$ converges to one faster for the low-$l$ modes as $\Omega_{ij}$ approaches zero. Similar argument can be applied to explain why in Fig.~\ref{fig:FW1} decreasing the included angle between the two partons will raise the FW spectrum at low $l$ and push its first acoustic peak to the high-$l$ end. The FW oscillation pattern thus can serve as a probe to the collimation of the visible particles and jet ancestral partons in each event.

\begin{figure}[h!]
	\centering
	\includegraphics[scale=0.29]{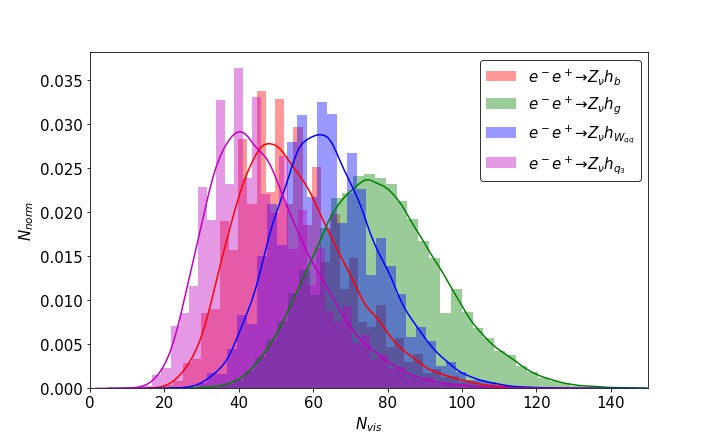}
	\includegraphics[scale=0.29]{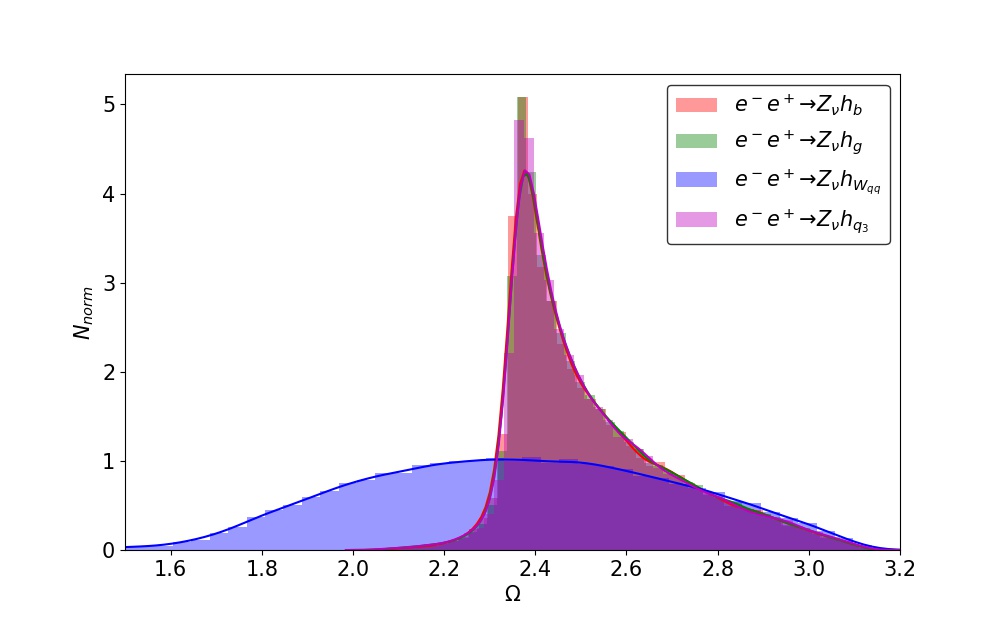}
	\caption{Distributions of the visible-particle number in each event (left) and the included angle between its two jet ancestral partons (right), for the two-jet samples. For the $Z_\nu h_{W_{qq}}$ events, the two partons are from the $W$ boson with a larger mass. The solid curves are generated by fitting.}
	\label{fig:nb_par}
\end{figure}

\item The FW spectrum picturizes the summation of the self- and inter-correlations of the visible particles (or the jet ancestral partons at leading order) at different angular scales. At $l=0$, the FW moment is given by $ H_{EE;0} = \frac{(\sum_i E_i)^2}{s}$. It represents the squared share of the visible energy among the total in each event. As is shown in Fig.~\ref{fig:FW2j}, the $ Z_\nu h_b$ events tend to have more missing energy, compared to the other ones. This can be explained by leptonic decays of bottom quarks.  At $l=1$, the FW moment is given by $H_{EE;1} = \frac{|\sum_i \vec{p}_i|^2}{s}$ (assuming all visible particles to be massless), with $\vec{p}_i$ being the particle three-momentum. $H_{EE;1}$ hence can serve as a measure of apparent momentum violation for the events at $e^-e^+$ colliders. For the two-parton system in Fig.~\ref{fig:FW1}, the $H_{EE;1}$ values show that its momentum is equal to zero at $\Omega =\pi$ and maximized at $\Omega =0$. Interestingly, a combination of $H_{EE;0}$ and $H_{EE;1}$ determines the visible and recoil mass of each event completely~\footnote{As a comparison, the CMB power spectrum for temperature fluctuation is physically less interesting for $l < 2$. At $l=0$ the moment is zero by definition. At $l=1$ the moment is dominated by the Doppler effect caused by the motion of the solar system w.r.t. the last scattering surface, which makes inseparable the cosmological dipole caused by large-scale perturbations. }. As was discussed above, the democracy effect arising from the self-correlation of the visible particles in each event and the collimation effect caused by their inter-correlation determine the profile of its FW spectrum. The FW spectra in Fig.~\ref{fig:FW2j} clearly demonstrate this. For example, the $Z_\nu h_g$ FW spectrum has the most-suppressed damping tail. This is mainly due to the stronger QCD radiation of gluons compared to those of quarks. As is shown in the left panel of Fig.~\ref{fig:nb_par}, this results in more visible particles in the $Z_\nu h_g$ events than the others. Notably, the flavor of jet ancestral partons can impact the showering also. From Fig.~\ref{fig:nb_par} and Fig.~\ref{fig:FW2j}, we can see that the $Z_\nu h_b$ events contain more visible particles (mainly due to bottom quark decays), compared to the $Z_\nu h_{q_{_3}}$, and hence their FW tail is suppressed more. Another example is the $Z_\nu h_{W_{qq}}$ spectrum. Each $Z_\nu h_{W_{qq}}$ event contains four jet ancestral quarks, in comparison to two of the $Z_\nu h_{g,b,q_{_3}}$ events. The inter-correlation at parton level thus becomes more involved in this case since it represents a collective effect of all possible parton pairings. To make the picture clear, we show the distributions of the included angle between the two (representative) jet ancestral partons in each event, in the right panel of Fig.~\ref{fig:nb_par}. For the $Z_\nu h_{W_{qq}}$ events, the two jet ancestral partons are selected to be from the $W$ boson with a larger invariant mass. These two partons tend to be harder, compared to the other two, and hence represent a more important contribution to $H_{EE;l}^{\rm inter}$ at parton level. This plot shows that the $Z_\nu h_{W_{qq}}$ distribution is much broader than the others. This is consistent with the observation in Fig.~\ref{fig:2jeventimg} that there is no clear pupil-like structure in the $Z_\nu h_{W_{qq}}$ cumulative Mollweide projection. More than that, different from the $Z_\nu h_{g,b,q_{_3}}$ parton pairs most of which have an included angle $\gtrsim \frac{3\pi}{4}$, a large portion of the $Z_\nu h_{W_{qq}}$ parton pairs prefer one $\lesssim \frac{3\pi}{4}$. This explains why at $l=3$ there is a convex in the $Z_\nu h_{W_{qq}}$ spectrum, and a concave instead for the others.

\item The two-jet events of each sample define a distribution w.r.t. $H_{EE;l}$, with the ones at $l=2$ being shown in the right panel of Fig.~\ref{fig:FW2j}. This reminds us that, unlike the CMB power spectrum, the $\langle H_{EE;l}\rangle$ spectrum is free from big sample variance, because of the large size of collider data. These distributions also manifest the order of the heights of the first acoustic peaks in the $\langle H_{EE;l}\rangle$ spectra, as is shown in the left panel of this figure. Notably, the $\langle H_{EE;l}\rangle$ spectra do not fully reflect the discrimination power of the FW moments. This power also relies on the event-distribution profiles of the data samples at each multipole.   

\end{itemize}

\begin{figure}[h!]
	\centering
	\includegraphics[scale=0.16]{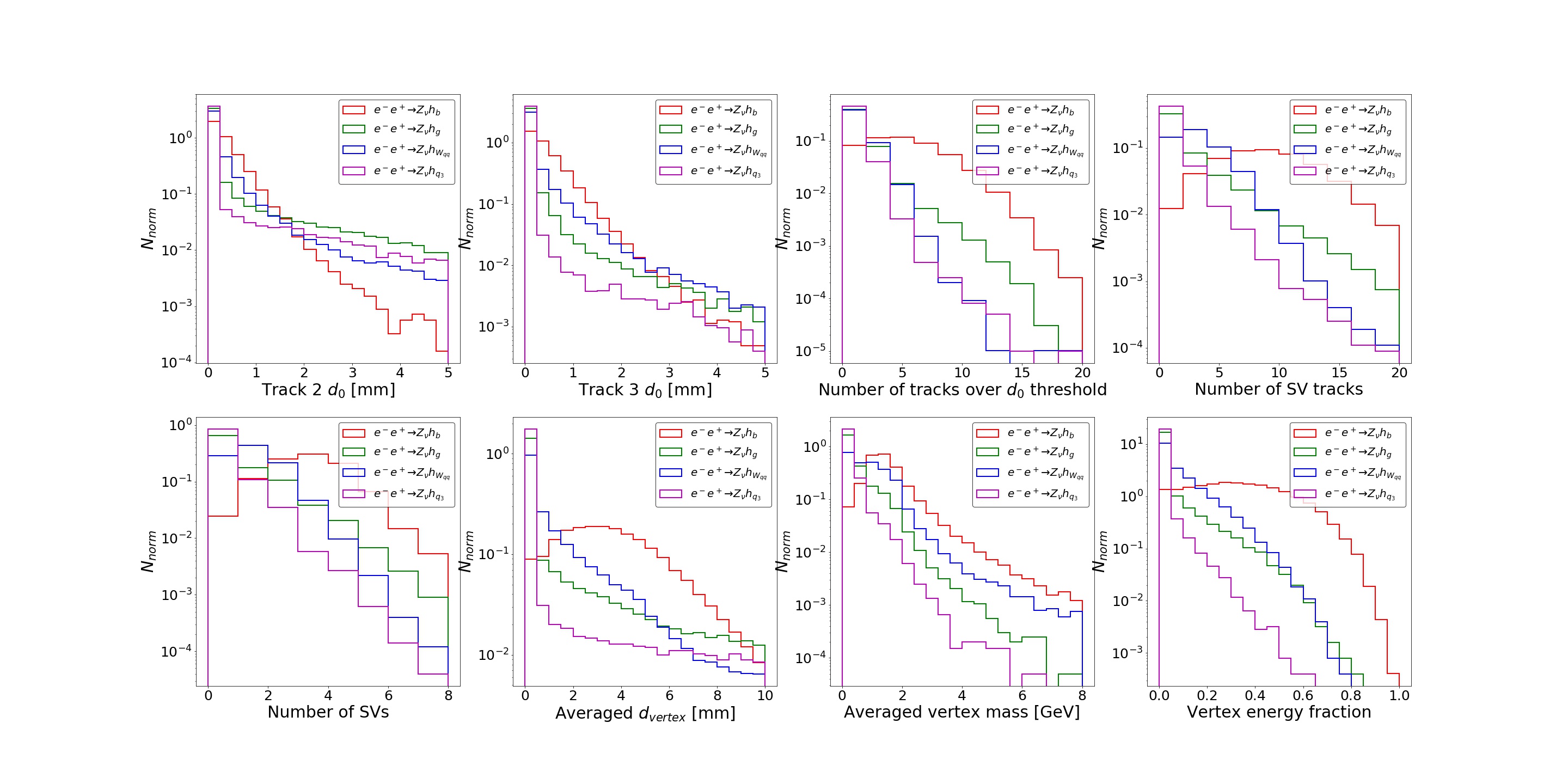}
	\caption{Normalized distributions of the (SV) track observables for the two-jet samples. Here the tracks in each event are sorted ascendingly with the impact parameter $d_0$. The $d_0$ threshold is set to be 0.5 mm. The vertex energy fraction is defined w.r.t. the total (PV + SV) track energy.}
	\label{fig:roc2ja}
\end{figure}
Another class of kinematic information arises from the tracks of the SV (defined by $d_{\rm vertex} \ge 10 \mu m$). As is well-known, the heavy-flavor quarks such as bottom tend to decay as a SV. This provides an important sign for recognizing these particles. Fig.~\ref{fig:roc2ja} displays the normalized distributions of a set of (SV) track observables for the two-jet samples (for simplicity, below we will not stress their ``SV'' nature). Here the tracks are defined at truth level with no detector smearing. These observables have been used for the DNN-based jet classification in~\cite{Guest:2016iqz}. As is expected, the $Z_\nu h_b$ events demonstrate highly distinguishable track features from the others, especially from the $Z_\nu h_{q_{_3}}$ events.  These features are shared to some extent by the $Z_\nu h_g$ and $Z_\nu h_{W_{qq}}$ events. This is largely because some heavy-flavor quarks such as charm quarks can be generated from gluon splitting and hadronic $W$ decay.

\begin{figure}[h!]
	\centering
	\includegraphics[scale=0.30]{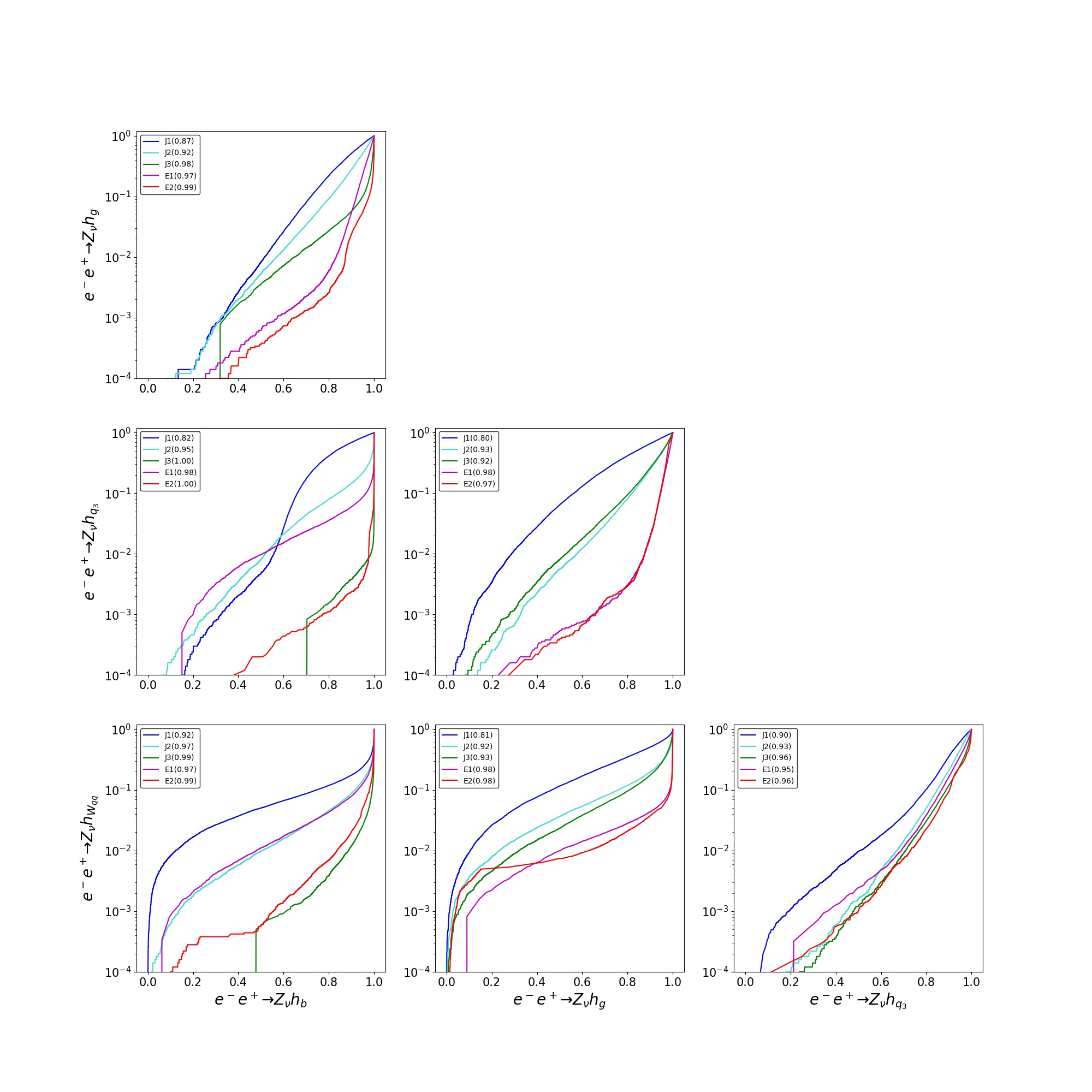}
	\caption{ROC curves and their AUC for the binary classifiers to distinguish between the four classes of two-jet events.}
	\label{fig:roc2jb}
\end{figure}
The receiver operating characteristic (ROC) curves and their area under the curve (AUC) for the binary classifiers to distinguish the four classes of two-jet events are presented in Fig.~\ref{fig:roc2jb}~\footnote{In this paper, the ROC curves are drawn as the acceptance of one classes of events against another one. So the ``AUC'' of each ROC curve is not really ``area under the curve'', but the area above the curve.}. In this figure, the blue and purple curves display the performance of J1 and E1 classifiers respectively. With the ML techniques of image recognition, the event-level classifiers are expected to be able to utilize the kinematic information to the greatest extent, upon the detector and image resolutions. Indeed, E1 classifiers yield an AUC universally bigger than that of J1 ones. The light-blue curves in this figure display the performance of J2 classifiers. They indicate that the FW moments of $H_{EE;l\leq 50}$ compensate for a large portion of the information lost at jet level. The AUC of J2 classifier even becomes comparable to that of E1 for the acceptance of the $Z_\nu h_b$ against the $Z_\nu h_{W_{qq}}$, due to this. It is also encouraging to see that the $Z_\nu h_b$ events can be tagged at $50\%$ level, with a faking rate of $1\%$ or below for the others, with no track observables being applied yet. For the case against the $Z_\nu h_{q_3}$, this mainly benefits from the flavor-related kinematics at event level such as multiplicity of the visible particles (see Fig.~\ref{fig:nb_par} and Fig.~\ref{fig:FW2j}). Despite these, the AUC gap between J1 and E1 classifiers is not fully addressed by $H_{EE;l\leq 50}$ in most cases. This may imply that the FW moments not included here or/and the multi-spectra carry part of the information lost at jet level. We will leave the relevant exploration to a future work. At last, the green and red curves display the performance of J3 and E2 classifiers. Not surprisingly, the incorporation of track observables yields a remarkable improvement to both jet-level and event-level classifiers in distinguishing the $Z_\nu h_b$ events from the others.

\subsection{Four-Jet Events}  
\label{subsec:4j}  

In the four-jet benchmark study, we will develop binary classifiers to distinguish between the four classes of $WW/ZZ$ events at $\sqrt{s} =240$ GeV, including: 
\begin{itemize} 

\item $e^-e^+ \to Z_qZ_q \to qqqq$;   

\item $e^-e^+ \to Z_qZ_b \to qqbb$; 

\item $e^-e^+ \to Z_bZ_b \to  bbbb$;

\item $e^-e^+ \to W_q W_q \to qqqq$. 

\end{itemize} 
Among these, the first three classes of events share the $ZZ$ production, but are differentiated by the flavor of their descendant partons. J2, J3, E1 and E2 classifiers are expected to be able to utilize the flavor-related event-level kinematics for their classification. The last class of events have different intermediate gauge bosons from those of the others. Their distinguishment may benefit additionally from the event-level kinematics manifesting the four momenta of the jet ancestral partons and even the nature of their parent gauge bosons. This is especially important for distinguishing between the $W_qW_q$ and $Z_qZ_q$ events, or probing anomalous triple-gauge couplings at $e^-e^+$ colliders.

\begin{figure}[h!]
\centering
\includegraphics[scale=.24]{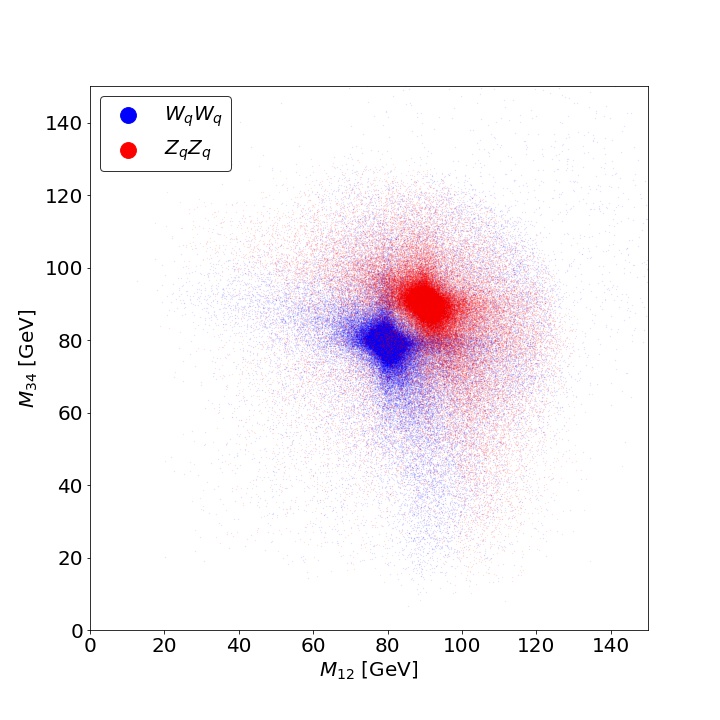}
\includegraphics[scale=.24]{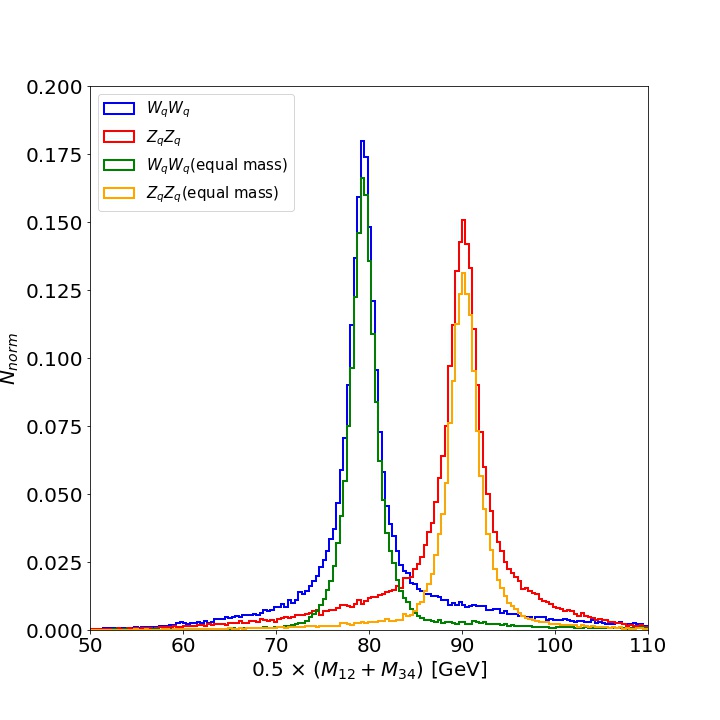}
\caption{Distributions of the reconstructed $W_qW_q$ and $Z_qZ_q$ events at the $m_{12} - m_{34}$ plane (left) and w.r.t. $(m_{12}+m_{34})/2$ (right), with a perfect detector. }
\label{fig:wwzztraditional}
\end{figure}
Actually, the $W_qW_q$ and $Z_qZ_q$ events have been applied to illustrate the problem of information deformation in jet clustering at CEPC~\cite{Zhu:2018ift}. In this study, these events were  reconstructed by minimizing the measure 
\begin{eqnarray}
\chi^2 = \frac{(m_{ab}-m_{X})^2+(m_{cd}-m_{X})^2}{\sigma_B^2}. 
\end{eqnarray}
Here $\{ab, cd\}$ runs over all possible jet pairings among the four, with $\{12, 34\}$ representing the optimal one, $X$ runs over $W$ and $Z$ bosons, and $\sigma_B$ denotes the standard deviation of the jet-pair invariant mass. In Fig.~\ref{fig:wwzztraditional} we show the distributions of the reconstructed $W_qW_q$ and $Z_qZ_q$ events at the $m_{12} - m_{34}$ plane and w.r.t. $(m_{12}+m_{34})/2$, with a perfect detector. Mainly due to the information deformation of jets, a good portion of these events are not well-reconstructed. This results in a separation of 50\% between these two classes of events. With a condition of mass equality  $|m_{12}-m_{34}|<10$GeV~\cite{Zhu:2018ift} being applied, this separation increases to 78\%, at the cost of losing 35\% $W_qW_q$ and 39\% $Z_qZ_q$ events. These results are consistent with the observations made in~\cite{Zhu:2018ift}. 

\begin{figure}[th!]
\centering
\includegraphics[scale=.3]{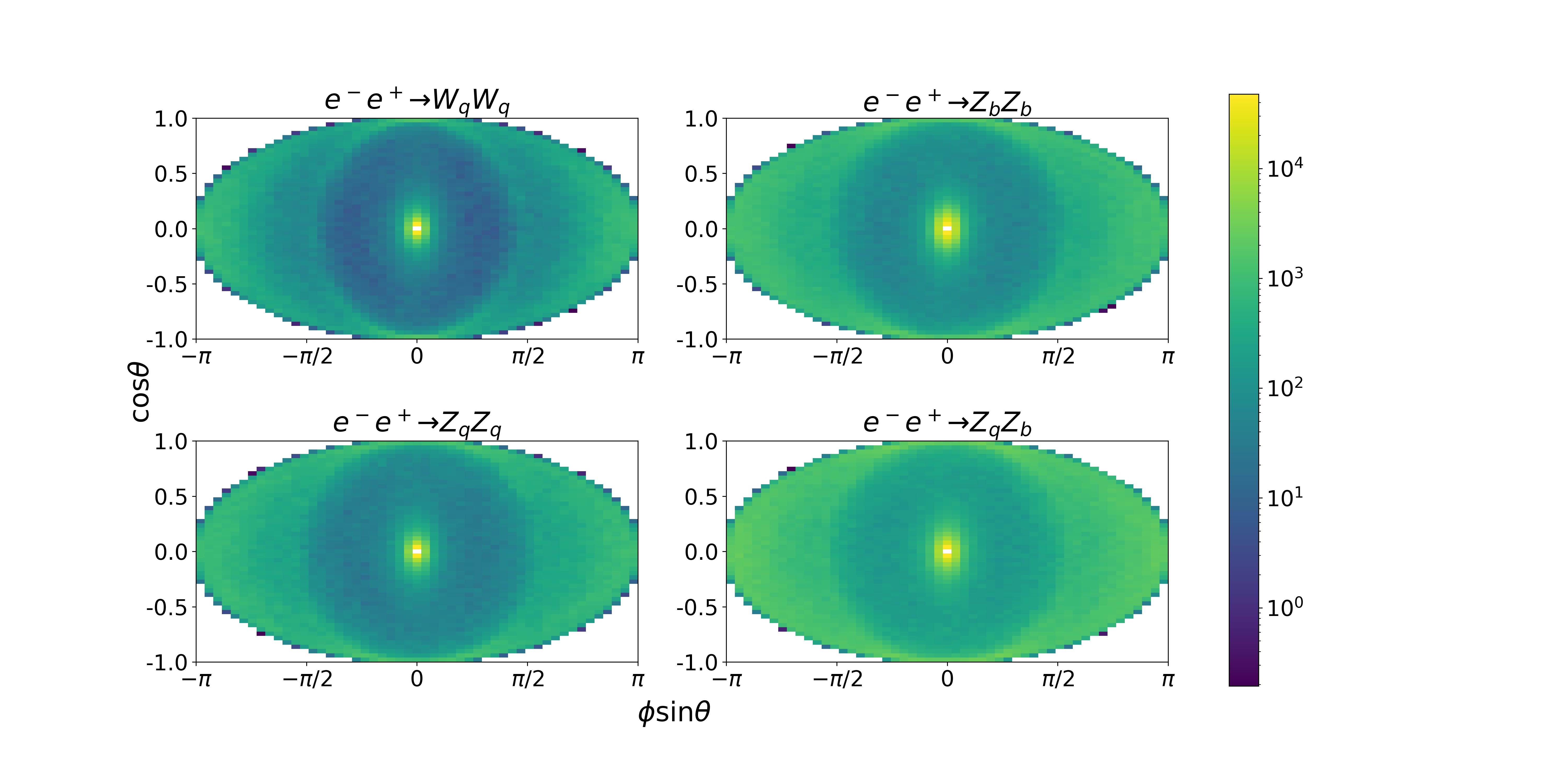}
\caption{Cumulative Mollweide projections of 10000 events:  $W_q W_q$ (upper-left), $Z_bZ_b$ (upper-right), $Z_qZ_q$ (bottom-left) and $Z_qZ_b$ (bottom-right), with the brightness of each cell scaling with the total energy (GeV) of the particle hits received. }
\label{fig:4jimg}
\end{figure}
\begin{figure}[h!]
	\centering
	\includegraphics[scale=0.295]{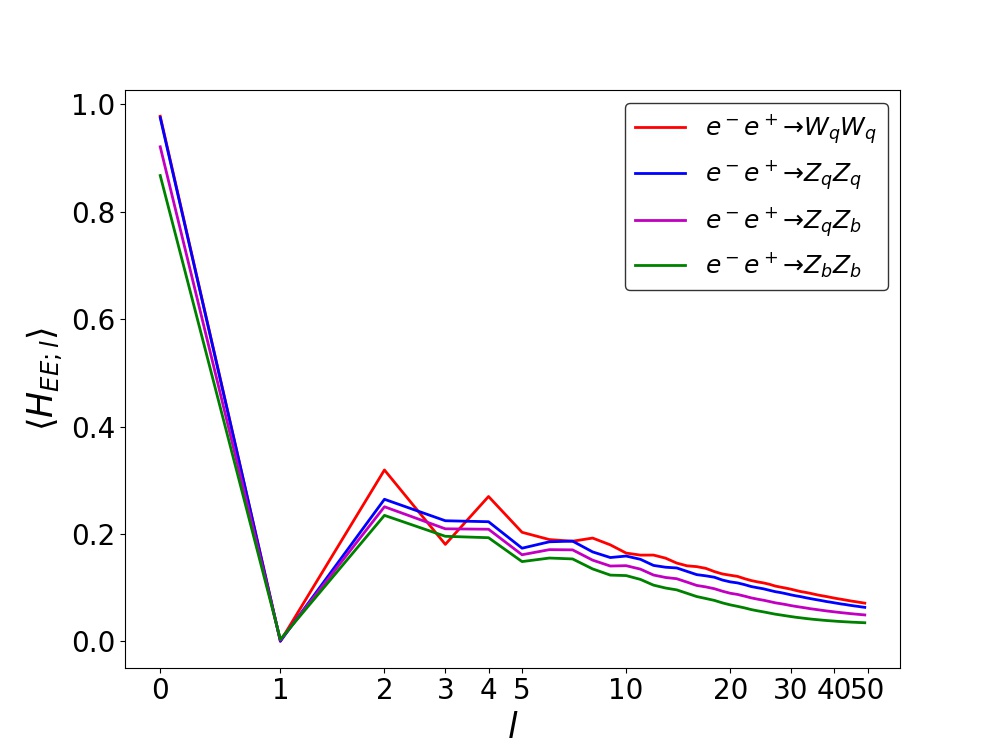}
	\includegraphics[scale=0.295]{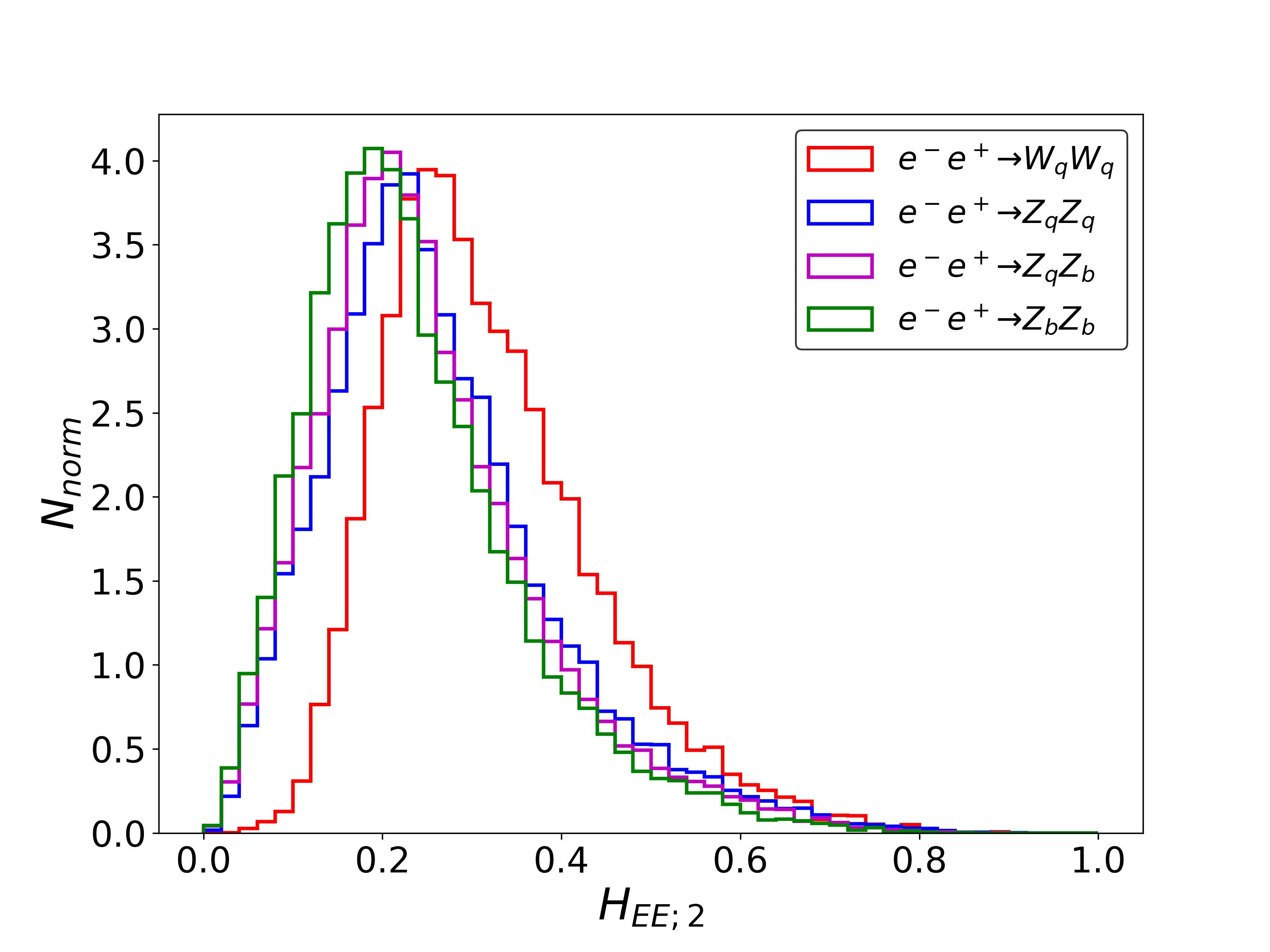}
	\caption{ FW spectra of $\langle H_{EE;l} \rangle$ (left) and event distributions of $H_{EE;2}$ (right) for the four-jet samples. }
	\label{fig:roc4ja}
\end{figure}
\begin{figure}[h!]
	\centering
	\includegraphics[scale=0.4]{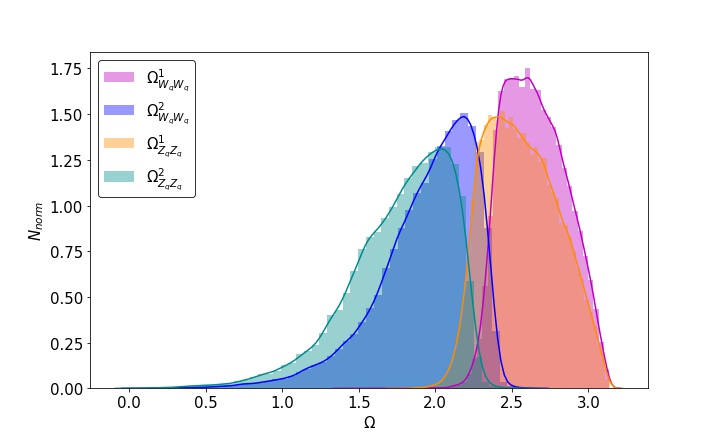}
	\caption{Distributions of the largest ($\Omega^1$) and the second largest ($\Omega^2$) included angles between the paired jet ancestral partons, for the $W_qW_q$ and $Z_qZ_q$ events. Here one of the paired partons is required to be the most energetic among the four, and another one to be from a different parent particle.}
	\label{fig:angle}
\end{figure}
We present the cumulative Mollweide projections of these four classes of four-jet events in Fig.~\ref{fig:4jimg}. All of them display a pupil-like structure clearly. 
For these events, each of them contains two intermediate gauge bosons moving oppositely. The $Z_qZ_q$ event is somewhat like a double copy of $Z_q$ in the $Z_\nu Z_q$ event, which explains why the pupil size of the $Z_qZ_q$ projection in Fig.~\ref{fig:4jimg} is comparable to that of the $Z_\nu Z_q$ projection in Fig.~\ref{fig:eventimg}. The $W_qW_q$ event has two more-boosted intermediate gauge bosons, due ot $m_W < m_Z$. Its cumulative Mollweide projection thus has a smaller pupil stucture, with its radius being smaller than $\frac{\pi}{2}$ at $\cos \theta=0$. In relation to this, one has a bigger chance to find the jet ancestral-parton pairs with an included angle close to $\pi$ in the $W_qW_q$ events, compared to the $Z_{q,b}Z_{q,b}$ ones. Most of such pairs are composed of the partons with different parent particles. Mainly due to this, in the $W_qW_q$ cumulative Mollweide projection, the region outside the pupil is approximately spilt to two: the dark one in the middle and the bright one close to the edge. We make this more explicit with Fig.~\ref{fig:angle}. It is exactly one to the partons comprising the magenda pair of $W_qW_q$ in this figure that results in the said bright region. Fig.~\ref{fig:roc4ja} displays the FW spectra of $\langle H_{EE;l} \rangle$ and the event distributions of $H_{EE;2}$ for the four-jet samples. As was discussed before, the FW oscillation pattern is a collective manifestation of the inter-correlations among all jet ancestral partons at leading order. Since the $W_qW_q$ events have a bigger chance to be found to have a parton pair with an obtuse included angle, it is not strange that the peaks in its FW spectrum are sharper than those of the $Z_{q,b}Z_{q,b}$ FW spectra.

\begin{figure}[h!]
	\centering
	\includegraphics[scale=0.3]{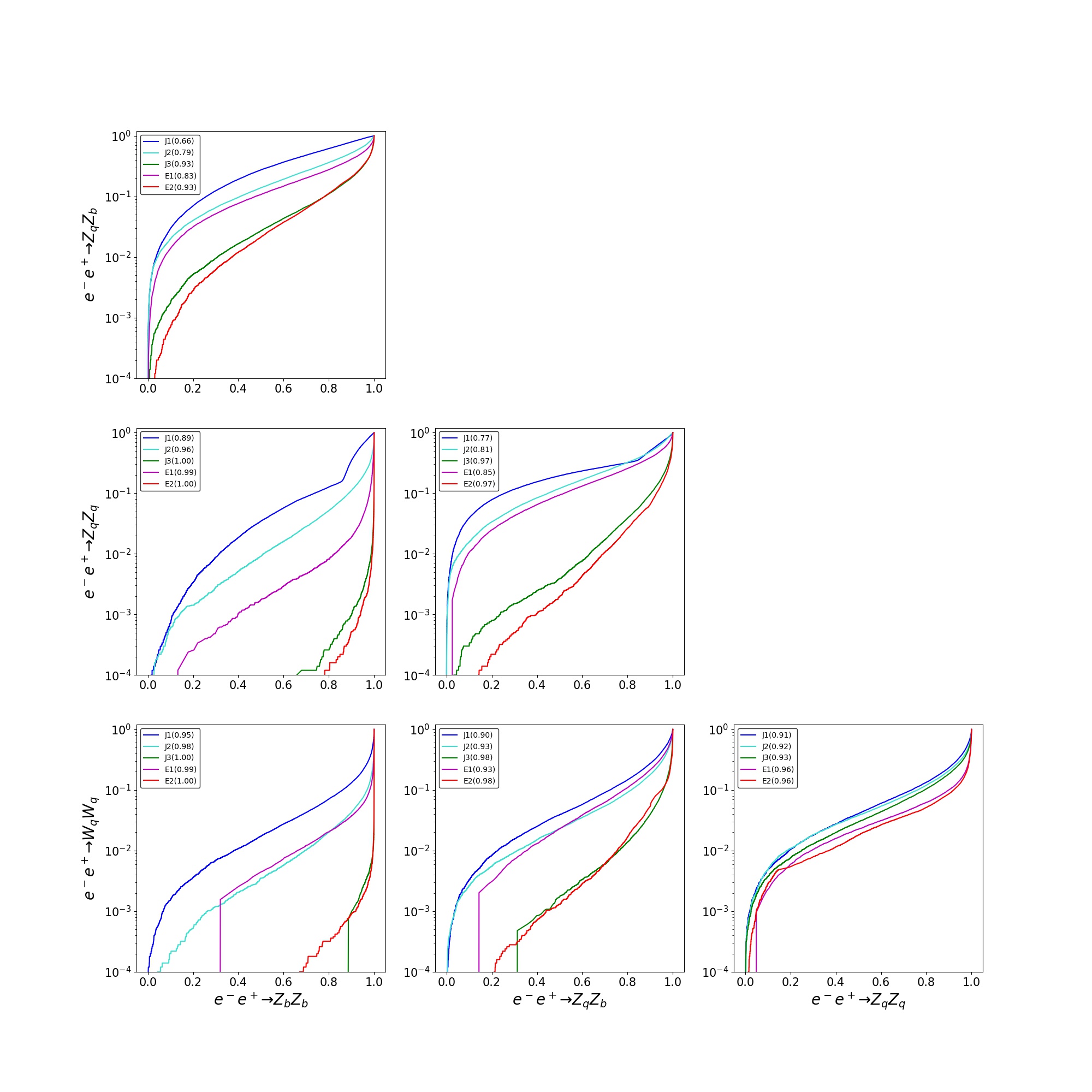}
	\caption{ROC curves and their AUC for the binary classifiers to distinguish between the four classes of four-jet events.}
	\label{fig:roc4jb}
\end{figure}
The ROC curves and their AUC for the binary classifiers to distinguish between the four classes of four-jet events are presented in Fig.~\ref{fig:roc4jb}. Similar to the two-jet case, E1 classifiers perform universally better than J1 ones. But, by incorporating the FW moments $H_{EE;l\leq 50}$, J2 classifiers greatly reduce their AUC gap in most cases. With the track information, J3 and E2 classifiers further improve the AUC values of J2 and E1. The extent is positively correlated with the difference of the bottom-quark number between the two classes of events to classify. Among these, the classifiers of the $Z_bZ_b$ against the others are especially informative. As is expected, its J1 classifiers perform best against the $W_qW_q$ and worst against the $Z_qZ_b$. Thus a space is created for the FW moments (or the event-level kinematics) and the track observables to play a non-trivial role in the latter case. Indeed, the $Z_bZ_b$ classifiers against the $Z_qZ_b$ gain the most from them among all constructions in Fig.~\ref{fig:roc4jb}, resulting in a great AUC improvement from 0.66 to 0.93 for both J3 and E2 ones.

\begin{figure}[h!]
	\centering
	  	\includegraphics[scale=.24]{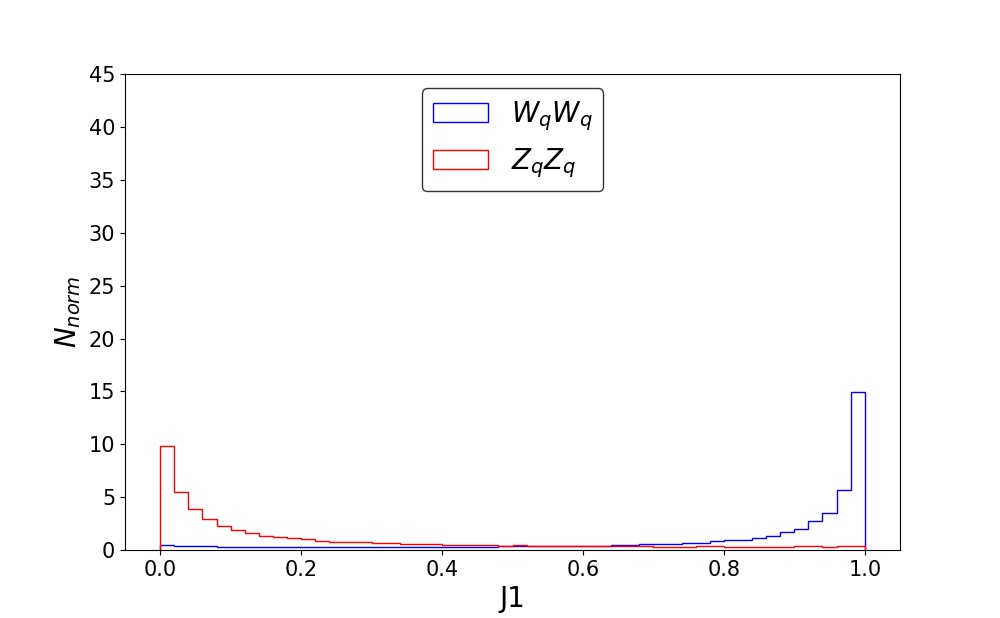} \\
			         \includegraphics[scale=.24]{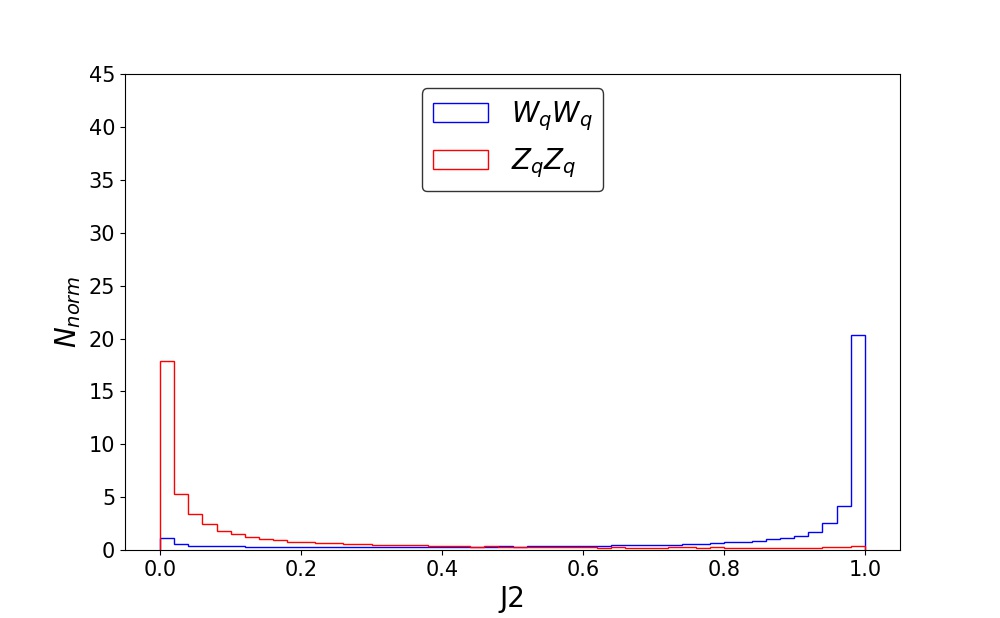}
		\includegraphics[scale=.24]{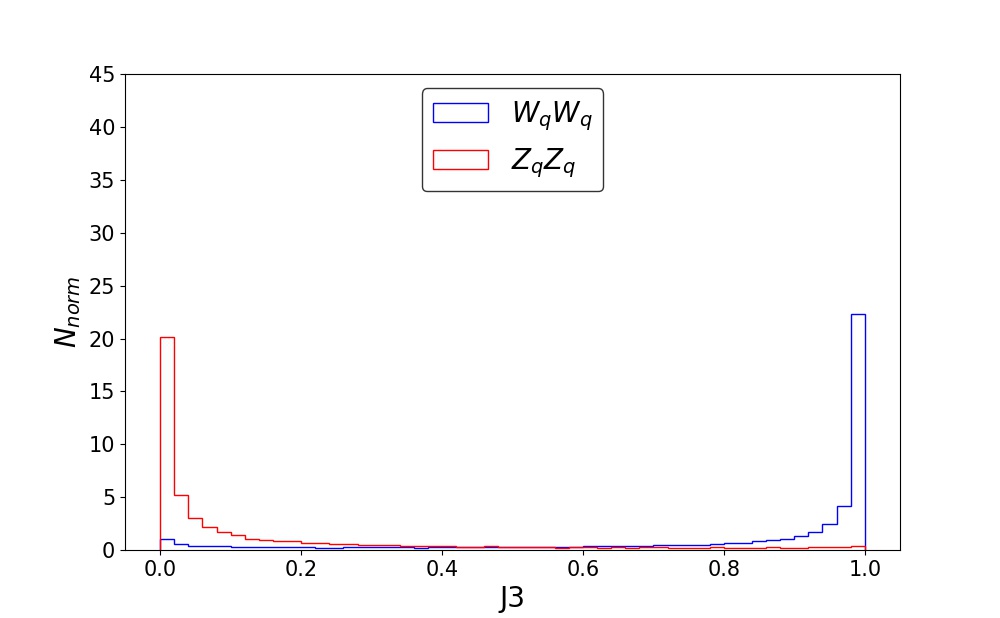} 
	         \includegraphics[scale=.24]{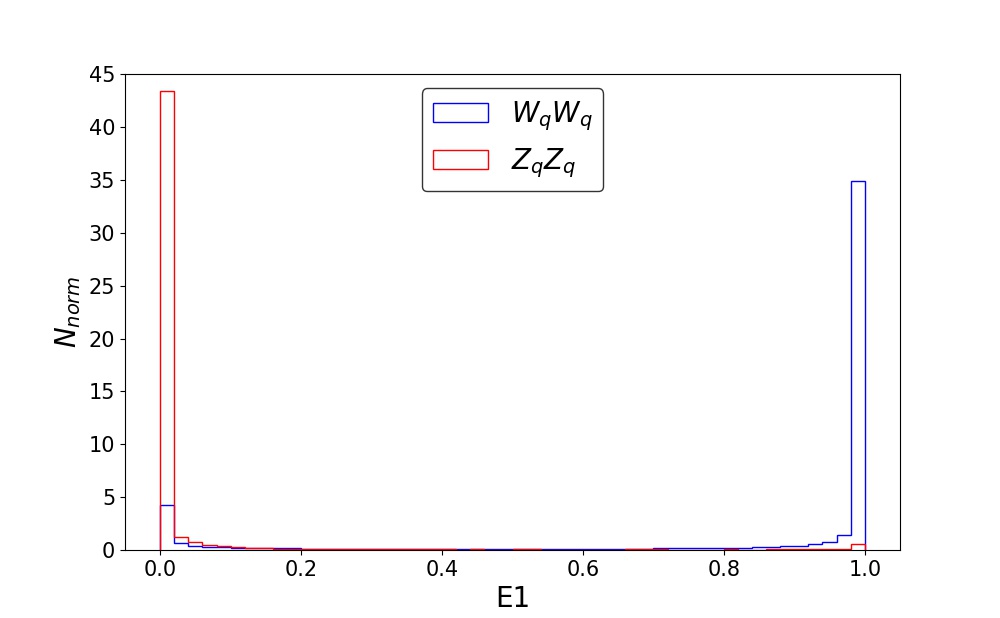}
		\includegraphics[scale=.24]{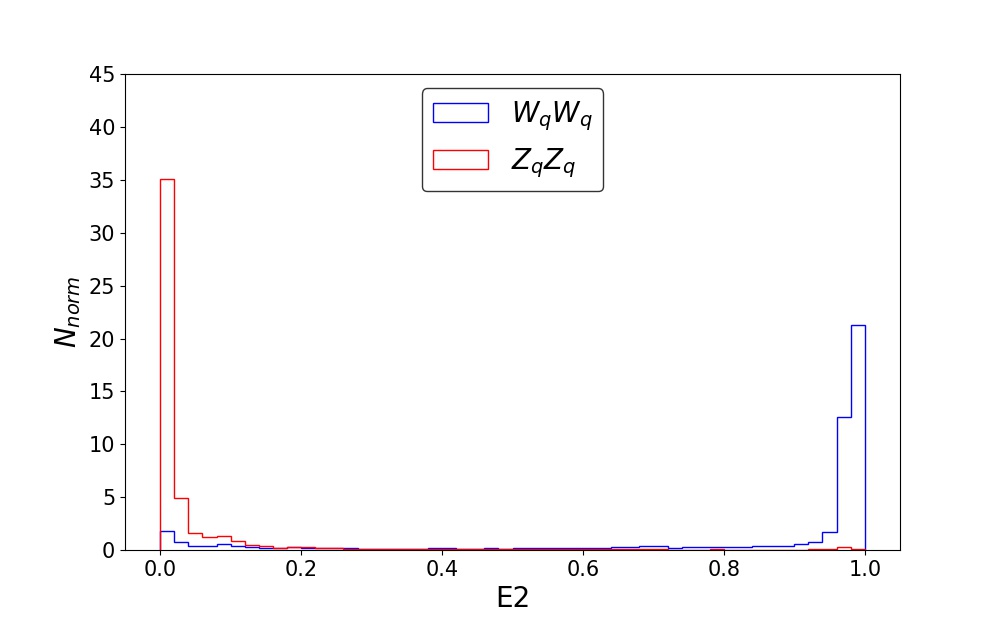}
  	 \caption{Response of the $W_qW_q$ and $Z_qZ_q$ events to J1, J2, J3, E1 and E2 classifiers.}
	\label{fig:4jetML}
  \end{figure}  
At last, let us take a look at the classification of the $W_qW_q$ and $Z_qZ_q$ events. We show the response of these two classes of events to J1, J2, J3, E1 and E2 classifiers in Fig.~\ref{fig:4jetML}. These classifiers all improve the event tagging accuracy to some extent, compared to the original analysis discussed above. But, E1 and E2 ones have a better performance than the others. They yield a separation of 70\% and 73\% respectively (J1: 53\%; J2: 56\%; J3: 59\%), without losing any events, in comparison to the original 50\%.

\section{Application: Higgs Decay Width ($\Gamma_h$)} 
\label{sec:Hmeasurement}

In this section we will apply the binary classifiers to measuring the SM $\Gamma_h$, one of the most important tasks at future $e^-e^+$ colliders, with the data of 5ab$^{-1}@240$GeV. 
Upon the strategy taken, this measurement may involve analyzing the events with two, four, and six jets. In this study, only the first two cases are relevant. 

\subsection{$\Gamma_h$ Measurement at $e^-e^+$ Colliders}
\label{subsec:Hwidth}

There exist multiple methods to measure the SM $\Gamma_h$ at $e^-e^+$ colliders. Here are several representative ones. 
\begin{itemize}

\item Method A. $\Gamma_h$ is measured using the relation 
\begin{equation}
\Gamma_h = \frac{\Gamma(h \to ZZ^*)}{\BR(h \to ZZ^*)}\propto \frac{[\sigma(Zh)]}{\BR(h \to ZZ^*)} = \frac{[\sigma(Zh)]^2}{[\sigma(Zh_Z)]} \ . \label{eq:ma}
\end{equation}
One needs to measure the quantities in the square brackets first for determining $\Gamma_h$. This method requires analyzing the $Zh$ data only, and hence is more straightforward compared to many other strategies. Its major drawback is that the $\sigma(Zh_Z)$ signal rate is small, while its irreducible backgrounds are relatively large.

\item Method B. $\Gamma_h$ is measured using the relation 
\begin{eqnarray}
 \Gamma_h = \frac{\Gamma(h \to WW^*)}{\BR(h \to WW^*)}    \propto \frac{\sigma(\nu\nu h)}{\BR(h \to WW^*)} 
 = \frac{[\sigma(\nu\nu h_b)] [\sigma(Zh)]^2}{[\sigma(Zh_b)][\sigma(Zh_W)]} \ .   \label{eq:mb}
\end{eqnarray}
This method utilizes the large signal rates of $\sigma(Zh_b)$ and $\sigma(Zh_W)$, and hence largely avoids the drawback of method A. 

\item Method C. $\Gamma_h$ is measured using the relation~\cite{Durig:2014lfa}  
\begin{equation}
\Gamma_h = \frac{\Gamma(h \to WW^*)}{\BR(h \to WW^*)}    \propto \frac{\sigma(\nu\nu h)}{\BR(h \to WW^*)} =\frac{[\sigma(\nu\nu h_b)]^2 [\sigma(Zh)]^2}{[\sigma(\nu\nu h_W)][\sigma(Zh_b)]^2} \ . \label{eq:mc}
\end{equation}
This method is similar to Method B, except that $\sigma(Zh_W)$, one of the key intermediate quantities to measure, is replaced with $\sigma(\nu\nu h_W)$. This method mainly benefits from the enhancement of the $\sigma(\nu\nu h)$ rate, as $\sqrt{s}$ increases. 

\end{itemize}
These three methods totally involve six intermediate quantities to measure: 
\begin{eqnarray}
\sigma(Zh), \ \sigma(Zh_Z), \ \sigma(\nu\nu h_b), \ \sigma(Zh_b), \  \sigma(Zh_W) \ {\rm and} \ \sigma(\nu\nu h_W) \ .    \label{list1}
\end{eqnarray}
With some of them, one can measure $\Gamma_h$ with a fourth method, $i.e.$,   
\begin{itemize}

\item Method D. $\Gamma_h$ is measured using the relation
\begin{eqnarray}
 \Gamma_h = \frac{\Gamma(h \to WW^*)}{\BR(h \to WW^*)}    \propto \frac{\sigma(\nu\nu h)}{\BR(h \to WW^*)} 
 = \frac{[\sigma(\nu\nu h_W)] [\sigma(Zh)]^2}{[\sigma(Zh_W)]^2} \ .   \label{eq:md}
\end{eqnarray}
This method shares the advantage of Method C, mainly benefitting from the enhancement of the $\sigma(\nu\nu h)$ rate at high $\sqrt{s}$.

\end{itemize}
With the relevant intermediate quantities being measured, one can calculate the precision of measuring $\Gamma_H$, using the formula of Gaussian statistics 
\begin{eqnarray}
\frac{\delta \Gamma_h}{\Gamma_h}  =  \sqrt{\sum_i \left (\frac{n_i \delta \mathcal{O}_i}{\mathcal{O}_i} \right)^2} \ .
\label{eq: precision}
\end{eqnarray}
Here $\mathcal{O}_i$ represents the intermediate quantities to measure in Eq.~(\ref{eq:ma} - \ref{eq:mc}, \ref{eq:md}), $\delta \mathcal{O}_i$ its absolute precision, and $n_i$ its power.

\begin{table}[th]
\centering
\resizebox{\textwidth}{!}{
\begin{tabular}{ccccccc}
\hline 
Measurements (\%) &  CEPC$_{240 (250)}$~\cite{An:2018dwb,CEPCStudyGroup:2018ghi}  & FCC$_{240}$~\cite{Abada:2019lih} & FCC$_{365}$~\cite{Abada:2019lih} & CILC$_{350}$~\cite{Abramowicz:2016zbo} & ILC$_{250}$~\cite{Durig:2014lfa,Ono:2013sea,LC-REP-2013-021,Li:2012taa}~ \\ 
\hline 
$\sigma(Zh)$ & 0.5 (0.5) & 0.5 & 0.9 & 1.6  & 2.8 \\ 
$\sigma(Zh_b)$ & 0.27 (0.26) & 0.3 & 0.5 & 0.86 & 1.2 \\ 
$\sigma(Zh_c)$ & 3.3 (3.1) & 2.2 & 3.5 & 14 & 8.3 \\ 
$\sigma(Zh_g)$  & 1.3 (1.2) & 1.9 & 6.5 & 6.1 &  7.0\\ 
$\sigma(Zh_W)$  & 1.0 (0.9) & 1.2 & 2.6 & 5.1  & 6.4 \\ 
 
$\sigma(Zh_Z)$  & 5.1 (4.9) & 4.4 & 12 & - & 19 \\ 
\hline 
$\sigma(\nu\nu h_b)$  & 3.2 (2.9) & 3.1 & 0.9 & 1.9 & 10.5 \\ 
$\sigma(\nu\nu h_c)$  & - & - & 10 & 26  & -\\ 
$\sigma(\nu\nu h_W)$  & - & - & 3.0 & - & - \\ 
\hline 
\end{tabular} 
}
\caption{Expected precisions of measuring $\mathcal{O}_i$ at $e^-e^+$ colliders. The CEPC$_{240}$ precisions are extrapolated from the CEPC$_{250}$ ones (inside the parentheses)~\cite{An:2018dwb,CEPCStudyGroup:2018ghi}. The ILC$_{250}$ results are based on its previous baseline luminosity (250~fb$^{-1}$).} 
\label{tab:obs}
\end{table}

\begin{table}[th] 
\resizebox{\textwidth}{!}{
\centering
\begin{tabular}{cccccc}
\hline 
$\Gamma_h$ (\%)  & CEPC$_{240 (250)}$~\cite{An:2018dwb,CEPCStudyGroup:2018ghi}  & FCC$_{240}$~\cite{Abada:2019lih} & FCC$_{240+365}$~\cite{Abada:2019lih}  & CLIC$_{350}$~\cite{Abramowicz:2016zbo} & ILC$_{250}$~\cite{Durig:2014lfa,Ono:2013sea,LC-REP-2013-021}~  \\ 
\hline 
Method A & 5.1 (5.0) & 4.5$^\ast$ & 4.2$^\ast$ & - & 20$^\ast$ \\
Method B & 3.5 (3.2)  & 3.5$^\ast$ & 1.7$^\ast$ & 6.7 & 13\\
Method C & - & - &  3.4$^\ast$ & - & - \\
Combined & 2.8 (2.7) & 2.7 & 1.3 & 6.7 & 11 \\ 
\hline 
\end{tabular} 
}
\caption{Expected precisions of measuring the SM $\Gamma_h$ at $e^-e^+$ colliders. The CEPC$_{240}$ precisions are extrapolated from the CEPC$_{250}$ ones (inside the parentheses)~\cite{An:2018dwb,CEPCStudyGroup:2018ghi}. The numbers marked with ``$^*$'' are derived from Table~\ref{tab:obs}, using Eq.~(\ref{eq: precision}).}
\label{tab:expfit}
\end{table}

The expected precisions of measuring $\mathcal{O}_i$ and $\Gamma_h$ in a variety of low-$\sqrt{s}$ operation scenarios at future $e^-e^+$ colliders are summarized in Table~\ref{tab:obs} and Table~\ref{tab:expfit}, respectively. Most of these analyses were pursued at jet level with a cut-based strategy. Method B provides a better precision of measuring $\Gamma_h$, compared to the other methods. Yet, to reduce the complication in Method B that the Higgs events with different decay modes serve as backgrounds mutually, one can instead measure the hadronic Higgs events in a relatively inclusive way. This idea can be also applied to Method C. We term these inclusive  methods as 
\begin{itemize}

\item Method B$'$. $\Gamma_h$ is measured using the relation 
\begin{eqnarray}
 \Gamma_h =\frac{\Gamma(h \to WW^*)}{\BR(h \to WW^*)}   \propto \frac{\sigma(\nu\nu h)}{\BR(h \to WW^*)}
 = \frac{[\sigma(\nu\nu h_h)] [\sigma(Zh)]^2}{[\sigma(Zh_h)][\sigma(Zh_W)]} \ . \label{eq:mbp}
\end{eqnarray}
Here $h_h$ denotes the inclusive two-body Higgs decays $h\to bb$, $cc$, $gg$ and $\tau\tau$. We exclude $h \to V_q V_q^*$ from $h_h$, to avoid a correlation between the $\sigma (Zh_h)$ and $\sigma(Zh_W)$ measurements. 

\item Method C$'$. $\Gamma_h$ is measured using the relation 
\begin{equation}
\Gamma_h = \frac{\Gamma(h \to WW^*)}{\BR(h \to WW^*)}    \propto \frac{\sigma(\nu\nu h)}{\BR(h \to WW^*)} =\frac{[\sigma(\nu\nu h_h)]^2 [\sigma(Zh)]^2}{[\sigma(\nu\nu h_W)][\sigma(Zh_h)]^2} \ . \label{eq:mcp}
\end{equation}

\end{itemize}
Then $\Gamma_h$ can be determined using either of Method A, B$'$, C$'$ and D and Eq.~(\ref{eq: precision}), with the new set of intermediate quantities 
\begin{eqnarray}
\sigma(Zh), \ \sigma(Zh_Z), \ \sigma(\nu\nu h_h), \ \sigma(Zh_h), \  \sigma(Zh_W) \ {\rm and} \ \sigma(\nu\nu h_W)    \label{list2}
\end{eqnarray} 
being measured.

We will take Method B$'$ in this study. Among the four intermediate quantities, $\sigma(Zh)$ can be measured with a precision of sub-percent level. $\sigma(Zh_h)$ is expected to be well-measured also, given the dominance of $\sigma(Zh_b)$ in its signal rate. As is shown in Table~\ref{tab:obs}, the precisions for both the $\sigma(Zh)$ and $\sigma(Zh_b)$ measurements are high. So we would expect the limitations for precisely measuring $\Gamma_h$ to mainly arise from the $\sigma(Zh_W)$ and $\sigma(\nu\nu h_h)$ measurements. Below we will focus on these two difficult cases. We will assume that all parameters relevant to their analyses have been precisely measured, and will not consider the impact of systematic errors.

\subsection{Measuring $\sigma(Zh) \BR(h \to WW^*)$}
\label{ssec:ww}

At CEPC, the measurement of $\sigma(Zh_W)$ was simulated with four decay modes of $Z$ boson~\cite{An:2018dwb}: $e^-e^+$, $\mu^-\mu^+$, $\nu\nu$ and $qq$. A combination of these yields a precision of 0.9\% at 250GeV (see Table~\ref{tab:obs}). The most important contribution arises from the two processes of $Z_\nu h_{W_{lq}}$ and $Z_\nu h_{W_{qq}}$ which give a combined precision of $1.5\%$~\cite{An:2018dwb}.  Below we will develop the binary classifiers for their measurements. 

\begin{table}
	\centering
\begin{tabular}{c|ccccccc}
			\hline \hline 
			Signal & & Backgrounds
		\\ \hline \hline
		  $Z_\nu h_{W_{lq}}$  & $W_lW_q$ & $Z_lZ_{q_{_5}}$ & $Z_\nu h_\tau$ & \\ \hline
		 $8.57\times 10^3$ & $2.41\times 10^5$  & $1.04\times 10^3$ &$3.22\times 10^3$  &  \\ \hline \hline
                     $Z_\nu h_{W_{qq}}$ & $Z_\nu Z_{q_{_5}}$ & $q_{_5}q_{_5} (\gamma)$ & $\gamma\gamma \to q_{_5} q_{_5}$  & $W_qW_q / Z_{q_{_5}} Z_{q_{_5}}$   \\ \hline 
			 $1.65\times 10^4$ & $5.61\times 10^4$  & $4.01\times 10^4$ & $4.41\times 10^2$  & $1.42\times 10^4$   \\ \hline
			 & $Z_\nu h_b$ & $Z_\nu h_c$ & $Z_\nu h_g$ & $Z_\nu h_{Z_{{q_{_5}}{q_{_5}}}}$ \\ \hline
			 &  $8.78\times 10^4$  & $4.71\times 10^3$   & $1.41\times 10^4$  &
		$2.10\times 10^3$  \\  \hline
			\hline
		\end{tabular}
		\caption{Numbers of the signal and its main background events after preselection, for measuring $\sigma(Z_\nu h_{W_{lq}})$ and $\sigma(Z_\nu h_{W_{qq}})$. We use Pythia8 to simulate the impact of initial state radiation and beamstrahlung for the $q_{_5}q_{_5}$ production. The $\tau$ decays of the $W/Z$ bosons are also incorporated in the $Z_\nu h_{W_{lq}}$ analysis, with $l= e, \mu$ and $\tau$.}
		\label{tab:hvvtotal}
\end{table}

To improve the training efficiency of these classifiers, we apply a set of preselection cuts first. For the $\sigma(Z_\nu h_{W_{lq}})$ analysis, we pass the events with one isolated lepton (either $e^-/e^+$ or $\mu^-/\mu^+$; $p_T > 10$GeV), and require the visible particles in each event including this lepton to have an invariant mass $\in[35, 125]$GeV, a recoil mass $\in [100, 200]$GeV, and a vector sum of transverse momentum $\in [10, 75]$GeV. The $W_lW_q$ events are then dominant in the backgrounds~\footnote{By requiring one hard isolated lepton, we notice that $\sigma (e \nu_e W_q)$, $i.e.$, the single-$W$ rate, is greatly suppressed compared to $\sigma(Z_\nu h_{W_{lq}})$. So we will not include these events in this analysis.}. The visible particles are subsequently clustered into two jets for the J1-, J2- and J3-based analyses. For the $\sigma(Z_\nu h_{W_{qq}})$ analysis, we veto the events with any isolated leptons and require the visible particles in each event to have an invariant mass $\in [100,150]$GeV, a recoil mass $\in [75,150]$GeV, and a vector sum of transverse momentum $\in [20, 80]$GeV. Different from the $Z_\nu h_{W_{lq}}$ case, both non-Higgs events such as $Z_\nu Z_{q_{_5}}$ and $q_{_5} q_{_5} (\gamma)$ and Higgs events including $Z_\nu h_{b}$ contribute to the backgrounds significantly. The visible particles are subsequently clustered into four jets for the jet-level analyses. The numbers of the signal and its main background events after preselection, for measuring $\sigma(Z_\nu h_{W_{lq}})$ and $\sigma(Z_\nu h_{W_{qq}})$, are summarized in Table~\ref{tab:hvvtotal}.

\begin{figure}[h!]
	\centering
	\includegraphics[scale=.29]{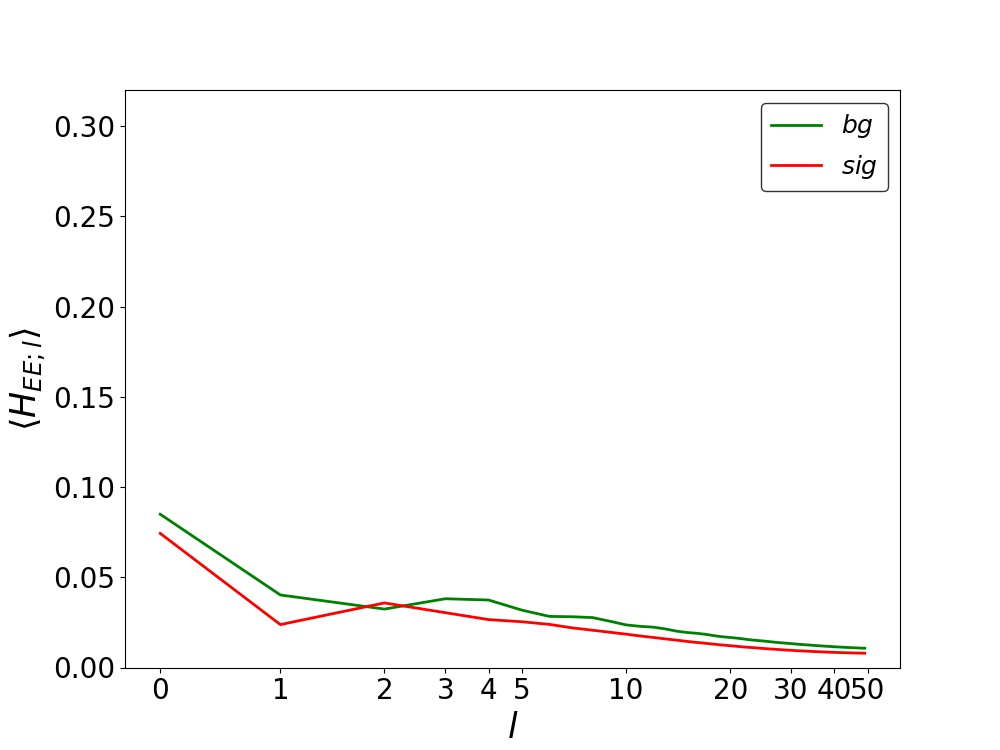}
	\includegraphics[scale=.29]{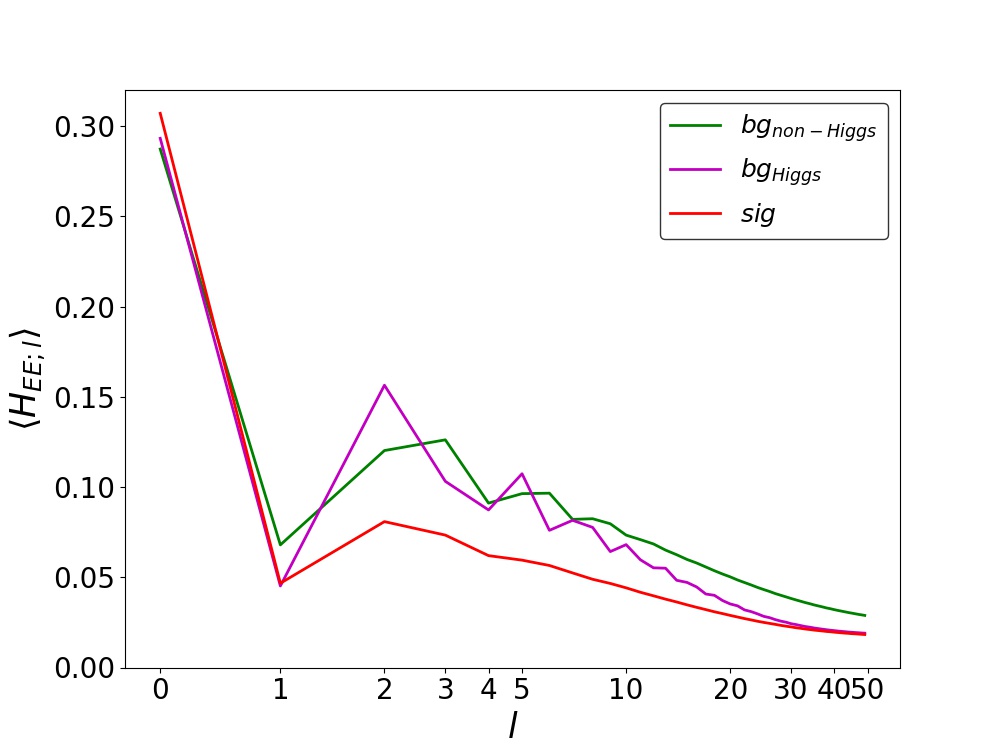}
		\includegraphics[scale=.29]{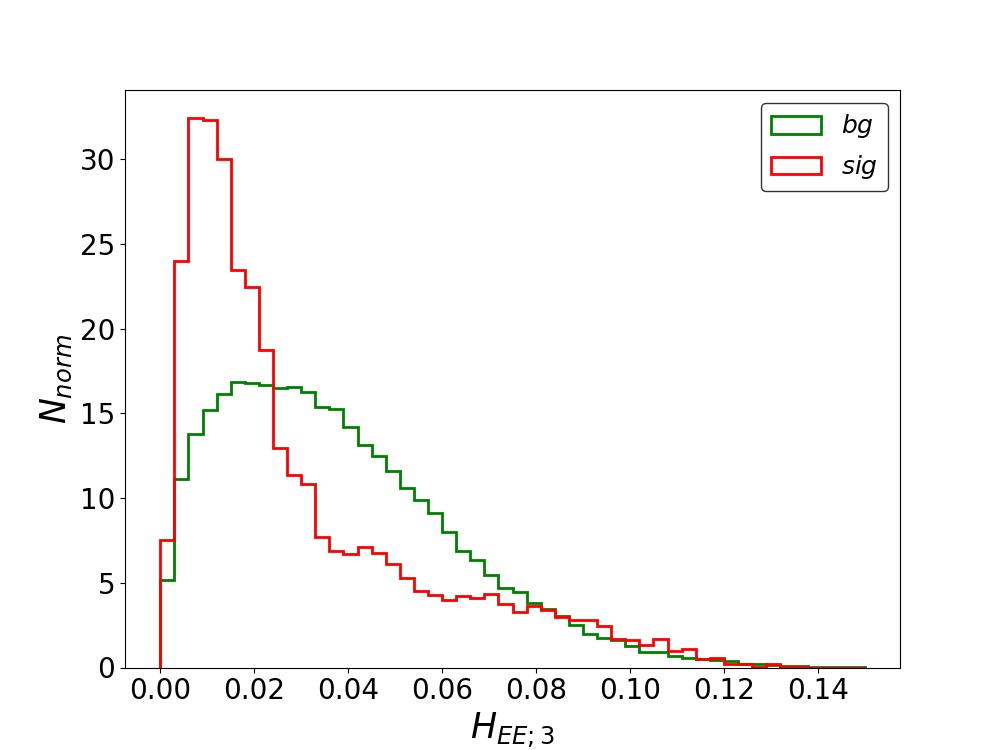}
	\includegraphics[scale=.29]{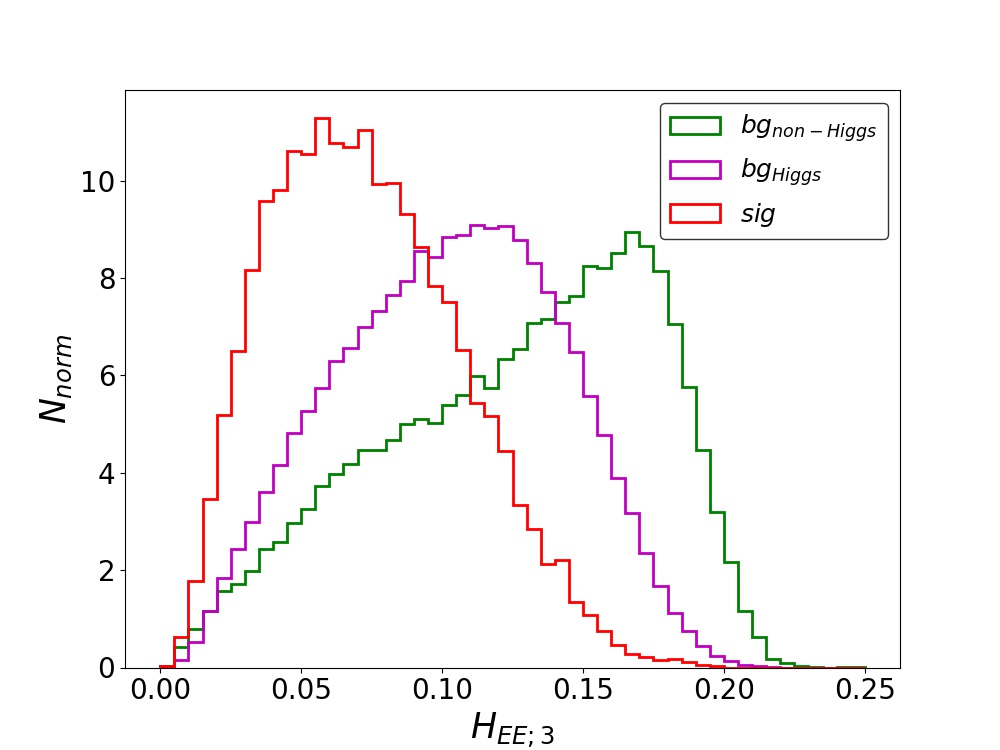}
	\caption{FW spectra of $\langle H_{EE;l} \rangle$ (upper) and event distributions of $H_{EE;3}$ (bottom) for the $Z_\nu h_{W_{lq}}$ (left) and $Z_\nu h_{W_{qq}}$ (right) and their respective background samples after preselection. In the left panels, the contributions from the isolated lepton have been excluded.}
	\label{fig:Zh2jFW}
\end{figure}

The FW spectra of $\langle H_{EE;l} \rangle$ for the $Z_\nu h_{W_{lq}}$ and $Z_\nu h_{W_{qq}}$, and their respective background samples are presented in the upper panels of Fig.~\ref{fig:Zh2jFW}. In both cases, the signal and backgrounds have close $\langle H_{EE;0} \rangle$ and $\langle H_{EE;1} \rangle$ values. This is not very surprising since these events are all preselected from the phase space in favor of the signal. Despite this, these $\langle H_{EE;l} \rangle$ spectra demonstrate a series of characteristic features which may assist distinguishing between the signal and backgrounds. Some of them are related to the discussions in Subsec.~\ref{subsec:2j}. In the upper-left panel, the signal and background spectra are characterized by a peak and a valley, respectively, at $l=2$. This indicates that the included angle between the two jet ancestral quarks of $W_l W_q$ is not far from $\frac{\pi}{2}$ (see Fig.~\ref{fig:FW1}), while the one of $Z_\nu h_{W_{lq}}$ tends to be wider. In the upper-right panel, the convex-concave structure of the FW spectra at $l=3$ indicates that the Higgs backgrounds tend to have a wider included angle between their jet ancestral partons, while the non-Higgs backgrounds favor a narrower one. The former case has been discussed before. The latter one can be understood also. For the $Z_\nu Z_{q_{_5}}$ events, without preselection this angle will be reduced to $\sim \frac{\pi}{2}$, while for the $q_{_5} q_{_5} (\gamma)$ events, most of the jet ancestral quarks are produced at $Z$ pole with the $Z$ boson being recoiled against their initial state radiation. 

As was discussed in Subsec.~\ref{subsec:2j}, the discrimination power of the FW moments also relies on the distribution profiles of the signal and background events at each multipole. For illustration, we present the ones of $H_{EE;3}$ for the $\sigma(Z_\nu h_{W_{lq}})$ and $\sigma(Z_\nu h_{W_{qq}})$ analyses in the bottom panels of Fig.~\ref{fig:Zh2jFW}. The relevant FW moments demonstrate certain features to distinguish between the signal and backgrounds in both cases. In the bottom-left panel, the signal distribution has a sharp peak at small $H_{EE;3}$, in comparison to a shape curved down for the background profile. In the bottom-right panel the non-Higgs background events tend to have a bigger $H_{EE;3}$ value for the relatively small included angle between their two jet ancestral partons.

\begin{figure}[h!]
	\centering
	\includegraphics[scale=.29]{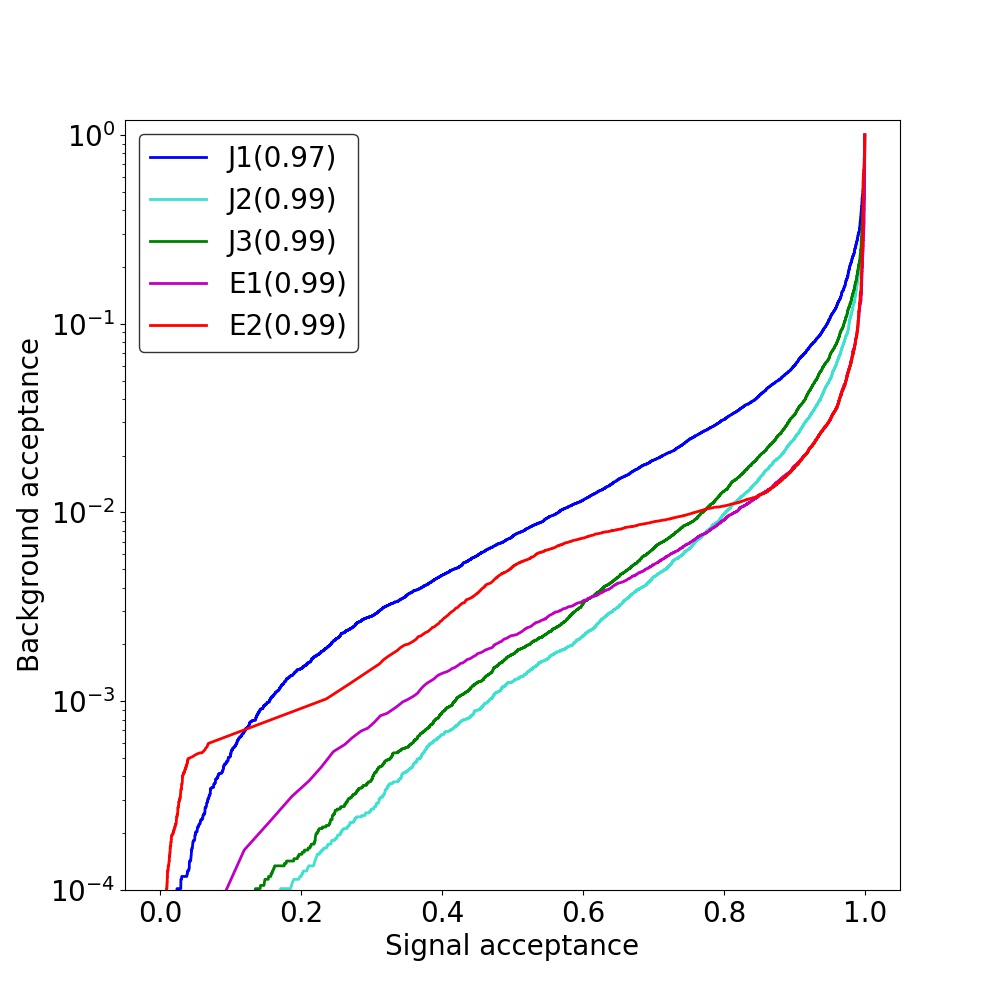}
	\includegraphics[scale=.29]{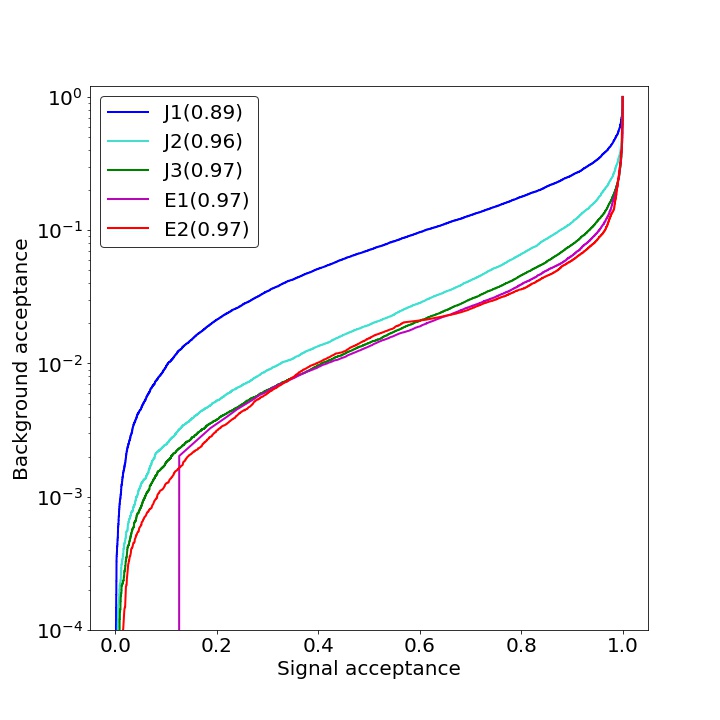}
	\caption{ROC curves and their AUC for the binary classifiers to distinguish the $Z_\nu h_{W_{lq}}$ (left) and $Z_\nu h_{W_{qq}}$ (right), from their respective background events.}
	\label{fig:Zh2jROC}
\end{figure}

Fig.~\ref{fig:Zh2jROC} displays the ROC curves and their AUC for the binary classifiers to distinguish the $Z_\nu h_{W_{lq}}$ and $Z_\nu h_{W_{qq}}$ from their respective backgrounds (the event responses to these classifiers are shown in Fig.~\ref{fig:eventvsML} in Appendix~\ref{app:A}). In both analyses, E1 classifiers yield an AUC bigger than that of J1 ones. Yet, by incorporating the FW moments $H_{EE;l\leq50}$, J2 classifiers fill up their gaps almost completely. The track observables are then applied to J3 and E2 classifiers, which improve the $\sigma(Z_\nu h_{W_{qq}})$ analysis slightly. In this case, the track information is useful in rejecting the $Z_\nu h_b$ events, the dominant Higgs background. A combination of these yields the same AUC values (up to $\mathcal O(10^{-2})$) for J2, J3, E1 and E2 classifiers of $\sigma(Z_\nu h_{W_{lq}})$ and for J3, E1 and E2 classifiers of $\sigma(Z_\nu h_{W_{qq}})$. 

The precisions of measuring $\sigma(Z_\nu h_{W_{lq}})$ and $\sigma(Z_\nu h_{W_{qq}})$  with these classifiers are summarized in Table~\ref{tab:MLHiggsWidth}. Not surprisingly, J2, J3, E1 and E2 classifiers result in comparable precisions in both analyses. Combining them allows the $\sigma(Zh_W)$ to be measured with a precision of $0.9\%$ at 240GeV. As a comparison, the CEPC baseline precision is $1.5\%$, which is achieved based on a cut-based analysis of the same channels, with the data of 5.6ab$^{-1}@250$GeV~\cite{An:2018dwb}.

\subsection{Measuring $\sigma(\nu\nu h)\BR(h\to {\rm hadrons})$}
\label{ssec:VBF}

In Method B, $\sigma(\nu\nu h_b)$ is a crucial intermediate quantity to measure. As is shown in Table~\ref{tab:obs}, the precision for its measurement is 3.2 (2.9)\% and 3.1\%, at CEPC$_{240(250)}$ and FCC$_{240}$, respectively. This comprises the main bottleneck for improving the precision of measuring $\Gamma_h$. In Method B$'$, we replace $\sigma(\nu\nu h_b)$ with an inclusive quantity $\sigma(\nu\nu h_h)$, for reducing unnecessary complexity in the analysis. This also brings an increasement of $\sim 22\%$ in the signal rate after event preselection. Notably, both methods suffer a subtlety caused by the interference between the signal of $\nu\nu h_{b,h}$ and its irreducible background $Z_\nu h_{b,h}$. To apply Eq.~(\ref{eq:mb}) and Eq.~(\ref{eq:md}), one needs to properly simulate this effect in the analysis. Yet, this was not explicitly implemented in~\cite{Fujii:2017vwa,An:2018dwb,Abramowicz:2016zbo,Abada:2019lih}. For the purpose of method comparison, we will tolerate this uncertainty below by simply neglecting it. We do not expect that such a treatment will qualitatively change the conclusions reached in this paper. 

\begin{table}[h!]
	\centering
	\resizebox{\textwidth}{!}{
	\begin{tabular}{c|cccccc}
	\hline \hline
	Signal & $\nu\nu h_b$ & $\nu\nu h_c$ & $\nu\nu h_g$ & $\nu\nu h_\tau$ &  \\ 	\hline 
	$1.51\times 10^4$ & $1.24\times 10^4$   & $6.43\times 10^2$   & $1.92\times 10^3$  & $1.50\times 10^2$   &  \\ 	\hline 
	Higgs backgrounds & $Z_\nu h_b$ & $Z_\nu h_c$ & $Z_\nu h_g$ & $Z_\nu h_\tau$  \\ \hline
		$1.39\times 10^5$  & $9.47\times 10^4$  & $5.08\times 10^3$   & $1.52\times 10^4$   & $1.06\times 10^3$  \\ \hline
	& $Z_\nu h_{V_{q_{_5} q_{_5} }}$ & $\nu\nu h_{V_{q_{_5}q_{_5}}}$ \\ \hline
 & $2.01\times 10^4$  & $2.51\times 10^3$   \\ 	\hline 	
	Non-Higgs backgrounds & $q_{_5}q_{_5} (\gamma)$/$\gamma\gamma \to q_5 q_5$ & $W_qW_q$ & $Z_{q_{_5}}Z_{q_{_5}}$ & $Z_\nu Z_{q_{_5}}$ &  \\ 	\hline 	
	$1.40\times 10^5$   & $6.79\times 10^4$/$2.81\times 10^3$  & $1.26\times 10^4$ & $6.61\times 10^2$   & $5.61\times 10^4$ & \\ 	\hline  \hline
	\end{tabular}
	}
	\caption{Numbers of the signal and its main background events after preselection, for measuring $\sigma(\nu \nu h_h )$. }
	\label{tab:VBF240total}
\end{table}

For training the classifiers efficiently, we preselect the events by requiring their visible particles to have total energy $\in[105, 155]$GeV, invariant mass $\in[100, 135]$GeV, recoil mass $\in [65, 135]$GeV, a vector sum of $p_T > 10$GeV and $p_z < 60$GeV, and vetoing the events with any isolated leptons. The visible particles in each event are subsequently clustered into two jets for the J1-, J2- and J3-based analyses. The numbers of the signal and its main background events after preselection are summarized in Table~\ref{tab:VBF240total}. Both Higgs events such as $Z_\nu h_h$ and non-Higgs events such as $Z_\nu Z_{q_{_5}}$ and $q_{_5}q_{_5}(\gamma)$ contribute to the backgrounds significantly.

\begin{figure}[h!]
	\centering
	\includegraphics[scale=.29]{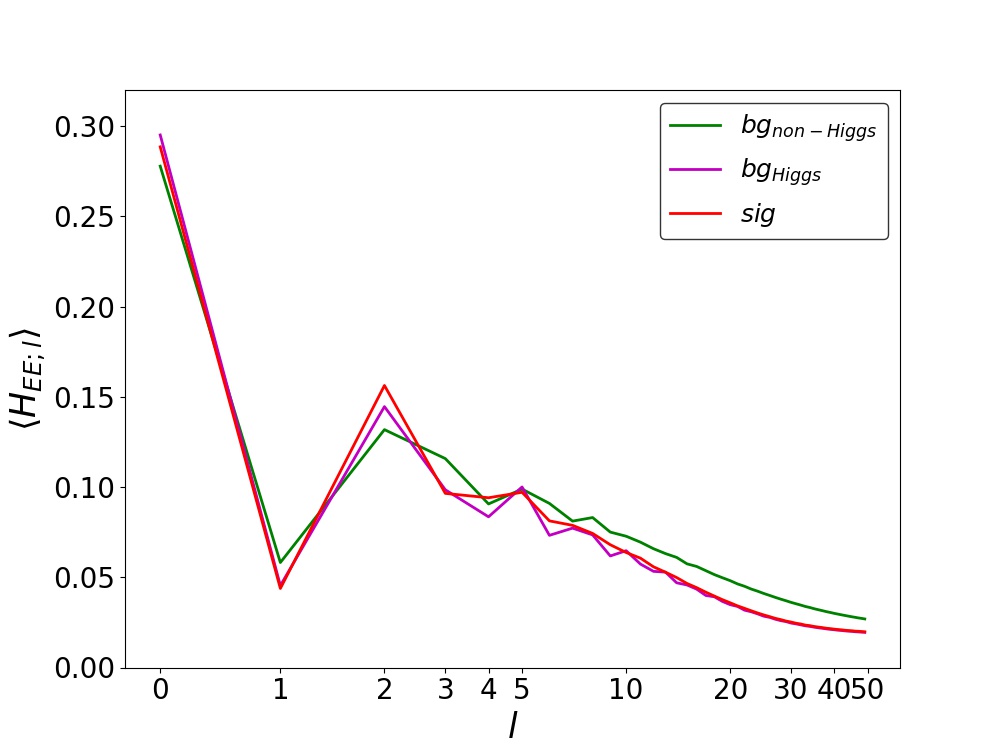}
		\includegraphics[scale=.29]{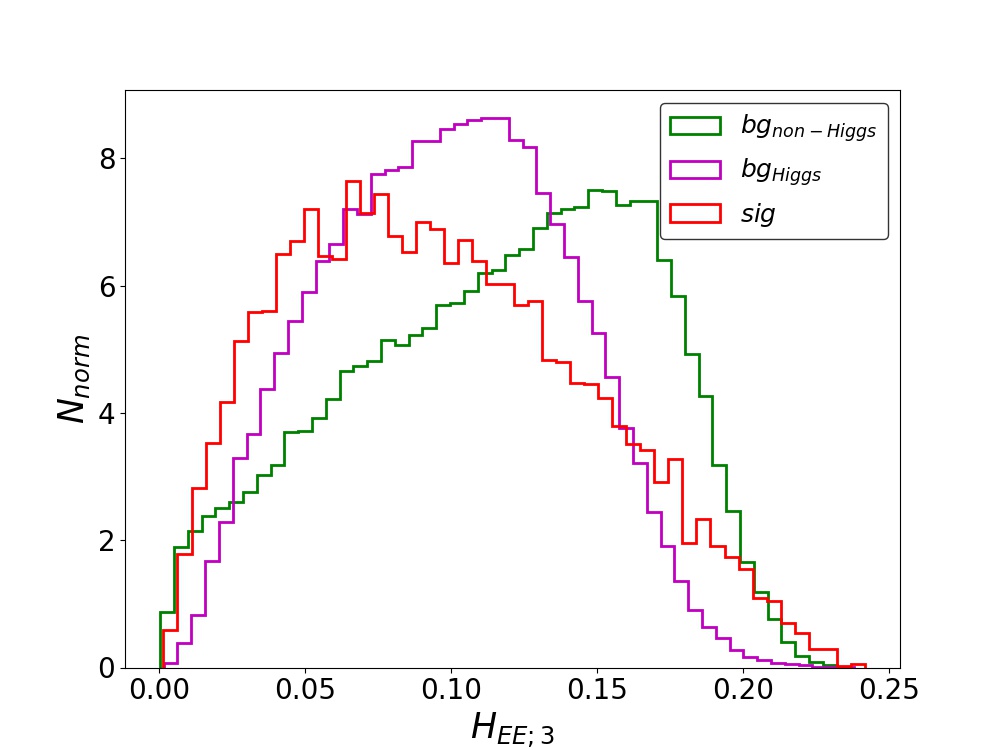}
	\caption{FW spectra of $\langle H_{EE;l} \rangle$ (left) and event distributions of $H_{EE;3}$ (right) for the $\nu\nu h_h$ and its background samples after preselection.}
	\label{fig:VBFFW}
\end{figure}

\begin{figure}[h!]
	\centering
	\includegraphics[scale=.29]{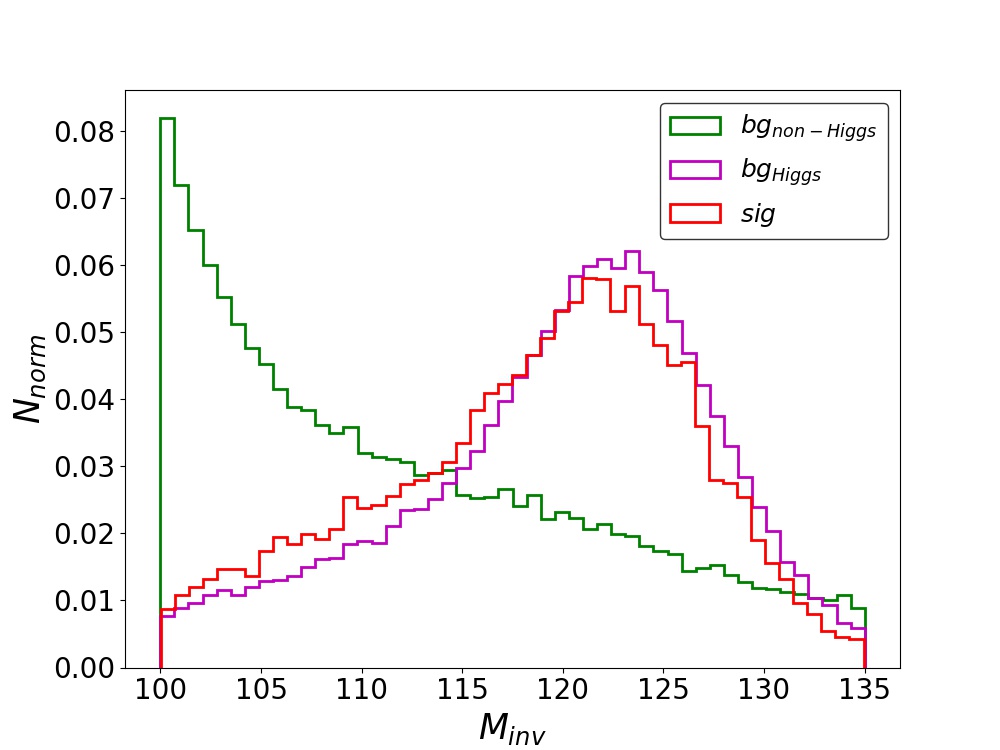}
		\includegraphics[scale=.29]{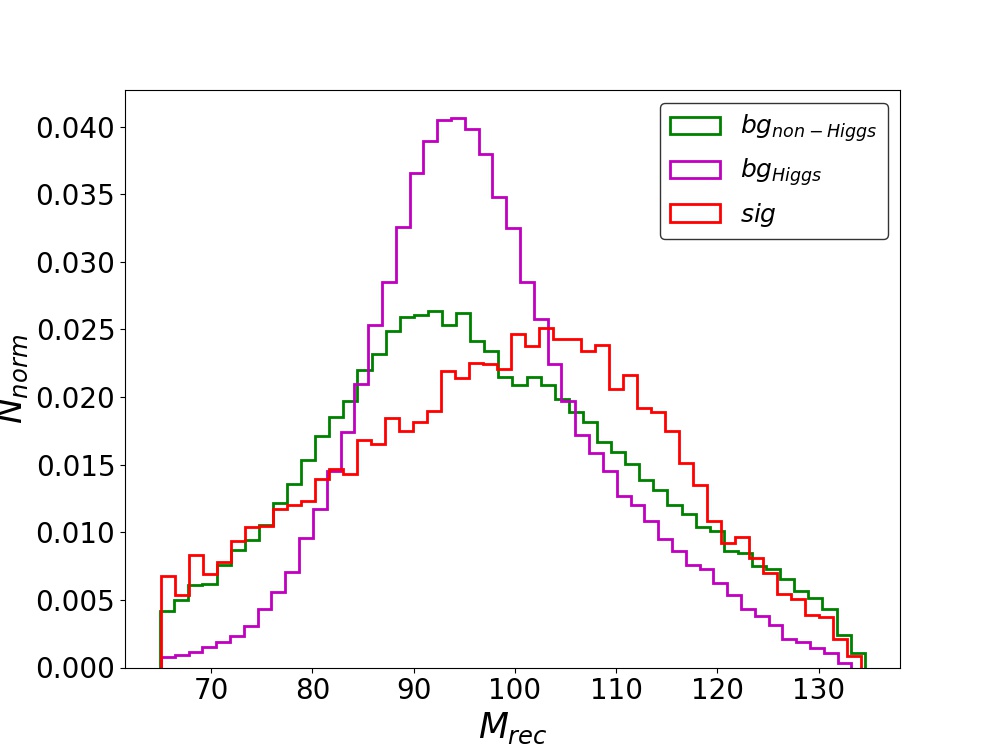}
	\caption{Event distributions of $M_{\rm inv}$ (left) and $M_{\rm rec}$ (right), defined in Eq.~(\ref{eq:mass}), for the $\nu\nu h_h$ and its background samples after preselection.}
	\label{fig:VBFFW1}
\end{figure}

Fig.~\ref{fig:VBFFW} displays the $\langle H_{EE;l}\rangle$ spectra and the event distributions of $H_{EE;3}$ for the $\nu\nu h_h$ and its background samples after preselection. Partly for preselection, these $\langle H_{EE;l}\rangle$ spectra are close to each other. But, as was stressed, the discrimination power of the FW moments also relies on the distribution profiles of the signal and background events at each multipole. For illustrating this, we show the event distributions of  the invariant and recoil masses of the visible particles in Fig.~\ref{fig:VBFFW1}. These two observables are determined by $H_{EE;0}$ and $H_{EE;1}$ completely, as was discussed in Subsec.~\ref{subsec:2j}, with the formulae given by 
\begin{eqnarray}
M_{\rm inv} = \sqrt{s (H_{EE;0} - H_{EE;1})};  \ \ \ \  M_{\rm rec} = \sqrt{s + M_{inv}^2 -  2 s \sqrt{H_{EE;0}}} \ .    
\label{eq:mass}
\end{eqnarray}
Both $M_{\rm inv}$ and $M_{\rm rec}$ (especially $M_{\rm inv}$) demonstrate certain discrimination power in Fig.~\ref{fig:VBFFW1}. In the left panel, the non-Higgs backgrounds tend to have a smaller $M_{\rm inv}$, compared to the others, since  $Z_\nu Z_{q_{_5}}$ and $q_{_5}q_{_5}(\gamma)$ are produced or mainly produced at $Z$ pole. In the right panel, as is expected, the Higgs background events are  accumulated near $Z$ pole. 

Beyond this, the $\langle H_{EE;l}\rangle$ spectra in Fig.~\ref{fig:VBFFW} demonstrates a series of distinguishable fine structures. One example is related to the FW moments at $l=2$ and $3$. The $\langle H_{EE;l}\rangle$ peaks at $l=2$ indicates that the $\nu\nu h_h$ and its main backgrounds favor an included angle bigger than $\frac{\pi}{2}$ for their jet ancestral partons. But, the ordering of the relevant samples w.r.t. $\langle H_{EE;3}\rangle$ (see both panels in Fig.~\ref{fig:VBFFW}) implies that the signal prefers the widest such angle while the non-Higgs backgrounds the smallest one. The kinematic information carried by these fine structures fails to be picked up by the main observables used for the traditional cut-based analysis of measuring $\sigma(\nu \nu h_b)$ (see, e.g.,~\cite{An:2018dwb}). These observables include $M_{inv}$, $M_{\rm rec}$ and the polar angle of Higgs boson~\footnote{By definition  the FW moments of $H_{EE;l}$ are not sensitive to event orientation in space, and hence are expected to be independent of the polar angle of Higgs boson. The information carried by the latter could be picked up by $H_{EE;l,m}$, $i.e.$, the FW moments at order $m$, in this CMB-like observable scheme. The  exploration regarding this is beyond the scope of this paper.}. The J2-, J3- and E1-, E2-based analyses of measuring $\sigma(\nu \nu h_h)$ thus may benefit a lot from such deformed or lost information at jet level and the overall information synergization.

\begin{figure}[h!]
	\centering
	\includegraphics[scale=.3]{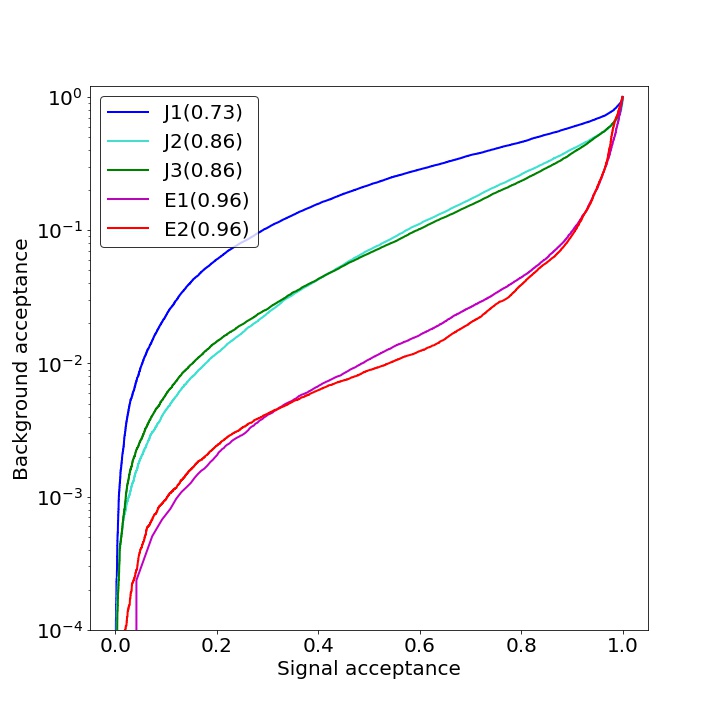}
	\caption{ROC curves and their AUC for the binary classifiers to distinguish the $\nu\nu h_h$ from its background events.}
	\label{fig:VBFROC}
\end{figure}

Fig.~\ref{fig:VBFROC} displays the ROC curves and their AUC for the binary classifiers to distinguish the $\nu\nu h_h$ from its backgrounds (the event responses to these classifiers are shown in Fig.~\ref{fig:eventvsML} in Appendix~\ref{app:A}). As is expected, E1 classifier demonstrates a better performance than that of J1. By including the FW moments $H_{EE;l\leq 50}$, J2 classifier yields a significant improvement to the AUC, $i.e.$, from 0.72 to 0.84. But, there is still an AUC gap between J2 and E1 classifiers. This gap could be filled up by the FW moments or/and multi-spectra which are not included in these jet-level analyses.

The precisions of measuring $\sigma(\nu\nu h_h)$ with these classifiers are summarized in Table~\ref{tab:MLHiggsWidth}. With the FW moments $H_{EE;l\leq 50}$, the jet-level precision is improved from 2.8\% (J1) to 1.8\% (J2) and 1.9\% (J3). The best precisions of $1.4\%$ and $1.3\%$ are achieved with the two event-level classifiers, $i.e.$, E1 and E2, respectively. These results indicate that, compared to the gain in the signal rate by replacing $\sigma(\nu\nu h_b)$ (Method B) with $\sigma(\nu\nu h_h)$ (Method B$'$), this measurement benefits more from synergizing the event-level information into the DNN-based analysis. A significant improvement to the precision is thus expected if these classifiers are applied to the exclusive measurement of $\sigma(\nu\nu h_b)$.

\subsection{Robustness against Detector Resolution}
\label{ssec:det}

The precisions of measuring the SM $\Gamma_h$ with Method B$'$, by applying J1, J2, J3, E1 and E2 classifiers to the data of 5ab$^{-1}@240$GeV, are summarized in Table~\ref{tab:MLHiggsWidth}. J1 classifiers yield a precision of 3.2\%. It is improved to 2.3\% by J2 and 1.9\% by E1, with the event-level information being incorporated. The track observables only have a slight impact for the measurements. The best outcome of $1.9\%$ improves the baseline precisions with Method B, $i.e.$, 3.5\% at both CEPC$_{240}$ and FCC$_{240}$, by a factor about 1.8.

\begin{table}[th]
\centering
\scalebox{1}{
\begin{tabular}{cccccc}
\hline 
\hline
Precision (\%) &  J1  & J2 & J3 & E1 & E2   \\ \hline 
$\sigma(Z_\nu h_{W_{lq}})$ &1.7 (1.6) & 1.4 (1.6)& 1.5 (1.6) & 1.5 (1.4) & 1.5 (1.4)   \\ 
$\sigma(Z_\nu h_{W_{qq}})$  & 1.6 (1.6)& 1.2 (1.2)& 1.1 (1.1) & 1.1 (1.1) & 1.1 (1.1)   \\ 
$\sigma(\nu\nu h_h)$ & 2.8 (2.7)& 1.8 (1.7)& 1.9 (1.8)& 1.4 (1.4)& 1.3 (1.3) \\  \hline 
$\Gamma_h$ & $3.2^{+0.9}_{-0.3}$ (3.1)& $2.3^{+0.7}_{-0.2}$ (2.2)& $2.3^{+0.7}_{-0.2}$ (2.3)& $1.9^{+0.5}_{-0.1}$ (1.9)& $1.9^{+0.4}_{-0.1}$ (1.9)  \\
\hline \hline
\end{tabular} 
}
\caption{Expected precisions of measuring $\Gamma_h$ with Method B$'$, by applying J1, J2, J3, E1 and E2 classifiers to the data of 5ab$^{-1}@240$GeV. In these analyses, a precision of 0.5\% for measuring $\sigma(Zh)$ and 0.3\% for measuring $\sigma(Zh_h)$ are assumed. The numbers in the parentheses represent the performance of these classifiers on the data processed with the FCC-ee IDEA template. The superscripts and subscripts for the numbers in last row denote the maximal changes of precision due to the variance of $\varepsilon$ from 1.0 to 2.0 and 0.1, respectively.} 
\label{tab:MLHiggsWidth}
\end{table}

\begin{figure} [th]
	\centering
	\includegraphics[scale=.3]{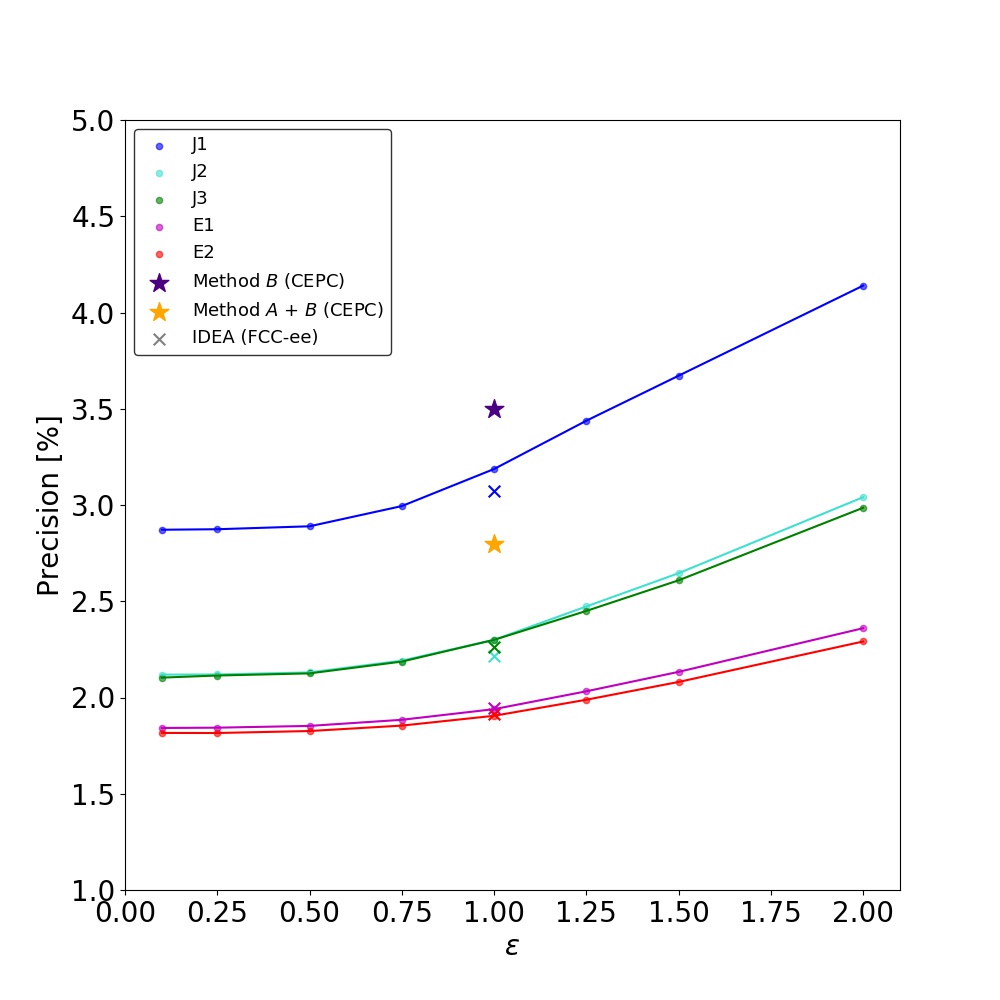}
\caption{Expected precisions of measuring $\Gamma_h$ versus detector energy/momentum resolution. $\varepsilon =1.0$ corresponds to the baseline resolution defined by the CEPC detector template~\cite{Ruan:2018yrh}. The purple and orange stars represent the precisions expected to be achieved with 5.6ab$^{-1}$ data at CEPC$_{240}$, using Method B and A+B, respectively~\cite{An:2018dwb,CEPCStudyGroup:2018ghi}. The colored crosses denote the precisions achieved by applying J1, J2, J3, E1 and E2 classifiers to the FCC-ee IDEA data at 240GeV. For the convenience of comparison, we place them at $\varepsilon =1.0$.}
\label{fig:detectoreffect}
\end{figure}

In these analyses, the detector effects are simulated with the built-in CEPC detector template~\cite{Ruan:2018yrh} of DELPHES3~\cite{deFavereau:2013fsa}. Such a specific choice naturally raises the question whether the classifiers developed are robust against the detector resolutions, including both energy/momentum and angular ones. To get some ideas about this, we take the following test. We first scale the energy/momentum resolution of track, ECAL, HCAL, electrons and muons defined in this template by a factor $\varepsilon$, then reprocess the data by varying its value from 0.1 to 2, and at last apply the classifiers developed at $\varepsilon =1$ for their testing. The precisions of measuring $\Gamma_h$ versus $\varepsilon$ are presented in Table~\ref{tab:MLHiggsWidth} and Fig.~\ref{fig:detectoreffect}. As one can see, the event-level classifiers are slightly more robust against the variation of $\varepsilon$ than the jet-level ones. For J1, J2 and J3 classifiers, the precision is improved by $8.7\% \sim 9.4\%$ as $\varepsilon$ decreases from its baseline value to 0.1, and degraded by $28\%\sim 30\%$ as $\varepsilon$ increases to 2. As a comparison, the precision is improved by $5.3\%$ and degraded by $21\% \sim 26\%$ in these two cases for E1 and E2 classifiers. But, no matter for which classifier, the conclusion reached on its performance has not been qualitatively changed. Additionally, in this study we incorporate the FW moments $H_{EE;l\leq 50}$ for J2 and J3 classifiers and image each event at a $50\times50$ evenly gridded $\theta-\phi$ plane in the E1- and E2-based analyses. Both setups are not fully synchronized with the baseline resolutions of the CEPC detector. The slightly worsening of the detector granularity or angular resolution thus could be absorbed by this uncertainty-tolerant space. A combination of these justifies the robustness of these classifiers against the detector resolutions. Beyond this, we test the robustness of these classifiers by simulating the testing data with the built-in FCC-ee IDEA template~\cite{IDEACard} of DELPHES3~\cite{deFavereau:2013fsa}. This FCC-ee detector benchmark differentiates itself from the CEPC one mainly by their calorimeter resolutions. The IDEA has an ECAL resolution of $0.11\sqrt{E}$ and an HCAL resolution of $0.30\sqrt{E}$ at leading order, in comparison to $0.20\sqrt{E}$ and $0.60\sqrt{E}$ of the CEPC detector template, respectvely. Also, the IDEA calorimeter granularity is higher than that of the CEPC HCAL and slightly lower than that of the CEPC ECAL. The remaining performance of these two detector templates are more or less comparable. The precisions achieved by applying J1, J2, J3, E1 and E2 classifiers to the FCC-ee IDEA data are presented in Table~\ref{tab:MLHiggsWidth} and Fig.~\ref{fig:detectoreffect} also. Compared to the CEPC results, they are better, but by no more than 5\%, for the $\Gamma_h$ measurements. 

It is noteworthy that these discussions never mean that better detector resolutions do not help much in improving the precision of measuring $\Gamma_h$. Recall, the detector granularity determines the highest multipoles of the FW moments which can be effectively applied to building J2 and J3 classifiers, and the largest pixels which can be legally used for constructing E1 and E2 ones. Also, the classifiers should be trained using the data processed at the new resolution benchmark. But, exploring this is beyond the scope of this study. 

Despite these tests, the uncertainty arising from the modeling of parton hadronization could impact the performance of the constructed classifiers (especially E1 and E2). It is known that the shower evolution is not simulated well in some cases such as gluon splitting to heavy flavors, fragmentation functions as $z\to 1$, etc. But, similar to J1, J2 and J3, which rely on infrared- and collinear-safe observables, E1 and E2 classifiers are based on the images which by definition are infrared- and collinear-safe. The finite resolution and particle-identity irrelevance of these event images grant them some level of immunity to hadronization details. Additionally, the impact of this uncertainty for the image-based classifiers have been studied in the context of jet classification~\cite{Komiske:2016rsd,deOliveira:2015xxd}, by analyzing the classifier performance with the data simulated by different event generators (Pythia, Herwig, Sherpa). One observation is that the light-quarks jets are less subject to this uncertainty compared with the gluon jets~\cite{Komiske:2016rsd}. This might be a sign of small impact of this uncertainty for the $\Gamma_h$ analysis, where mainly the quark jets are relevant. At last, we would bring it to the attention that the plenty of clean hadronic events produced at future $e^-e^+$ colliders may allow precisely matching data and simulation, and hence yield a suppression to this uncertainty.

\section{Summary and Discussion}
\label{sec:summary}

The $e^-e^+$ colliders, because of their clean QCD environment and absence of pileups, play a leading role in advancing the precision frontier in particle physics. One such machine of next generation is expected to push the precisions of measuring Higgs and electroweak physics up to an unprecedented level. Yet, due to the dominance of the hadronic events in data, many of the baseline precisions presented in documents are based on jet-level analysis and hence are limited by the information deformation and loss in jet clustering. We showed that this difficulty can be well-addressed by synergizing the event-level information into the DNN-based data analysis. In relation to this, we introduced a CMB-like observable scheme, where the event-level kinematics is encoded as the FW moments at leading order and multi-spectra at higher orders. Then we developed a series of jet-level (w/ and w/o the FW moments) and event-level binary classifiers, and analyzed their sensitivity performance comparatively with the two-jet and four-jet events. The general conclusion is: the event-level classifiers perform better compared to the jet-level ones; but, incorporating the FW moments into the jet-level classifiers can significantly reduce the performance gap between them.  As an application of such classifiers, we analyzed the precision of measuring the SM $\Gamma_h$ at $e^-e^+$ colliders with the data of 5ab$^{-1}@$240GeV. The precisions obtained are significantly better than the baseline ones. 

Yet, this is just an initial effort. We can immediately see several directions for future explorations. First of all, we showed that with the classifiers developed we are able to measure $\Gamma_h$ with a precision of $1.9\%$ (Method B$'$, E1 and E2 classifiers), improving the baseline ones (Method B) by a factor about 1.8. One natural question is if this precision can be pushed to sub percent level in a similar collider operation scenario. After all, the precision of measuring the SM $\Gamma_h$ is one of the most important indices to measure the physics capability of a future Higgs factory. This could be possible. As was discussed in Sec.~\ref{sec:Hmeasurement}, $\sigma(Zh_W)$ is one of the main intermediate quantities to determine the precision of measuring $\Gamma_h$ in both Method B and B$'$. In the CEPC analysis, the decay modes of $Z\to qq, l^+l^-,\nu\nu$ are combined for the $\sigma(Zh_W)$ measurement~\cite{An:2018dwb}. But, we exclusively focused on the $Z\to \nu\nu$ mode in this study. Also, we have assumed the precisions obtained from the cut-based analyses for measuring $\sigma(Zh)$ and $\sigma (Z h_h)$. A more complete analysis is thus necessary and important. Beyond that, the SM $\Gamma_h$ can be determined with four different methods at $e^-e^+$ colliders, with the set of six intermediate quantities in Eq.~(\ref{list2}) being measured. One may consider combining these methods, to yield an overall precision. But, as a reminder, one needs to take into account the systematic errors properly, before a firm statement regarding this can be made. 

Secondly, this effort opens a new angle to evaluate the physics capability of the future $e^-e^+$ colliders. We expect the developed strategies to be applied to many other hadronic measurements. Such measurements include Higgs couplings and CP properties, electroweak precision parameters, flavor physics, top physics, QCD parameters, etc. The applications can be even extended to new physics searches via high-$\sqrt{s}$ $e^-e^+$ collisions. To fully evaluate the collider capability, one needs to pursue a comprehensive study on these aspects. 

Thirdly, we expect that with the CMB-like observable scheme the kinematic information lost at jet level can be systematically reconstructed. Here we tested only to what extent the FW moments of energy, as part of the leading-order CMB-like observables, can compensate for that. We have observed that the incorporation of these FW moments can greatly reduce the performance gap between the jet-level and event-level classifiers in a general context, but can not eliminate completely. It is thus interesting to explore if the existing gap can be filled by the FW moments not included in this study and the multi-spectra. Taking a step further, we can leave the jet information out, and study comparatively the classifier based on the CMB-like observables only and the one with the techniques of image recognition. This will allow us to test the (approximate) completeness of this CMB-like observable scheme, and dissect the underlying physics of the event-level kinematics. 

Last but not least, although the CMB-like observable scheme was introduced for analyzing the data at $e^-e^+$ colliders, its application can be extended to, e.g., $ep$ colliders,  LHC and even future hadron colliders. But, the FW moments and the multi-spectra could be strongly smeared at such machines because of the four-momentum anisotropy of their events, contamination of pileups, etc. If these problems can be well addressed, we would expect the CMB-like observable scheme to be a powerful tool as well in these collider scenarios. We will leave this study and the others to a future work.

\section*{Acknowledgements}

We would thank Michelangelo Mangano and Manqi Ruan greatly for informative communications on the FCC-ee and CEPC analyses and for valuable comments on this manuscript. We would also thank Spencer Chang, Shirley Ho, Xuhui Jiang, Gang Li and David Shih for useful discussions. This research was supported partly by the General Research Fund (GRF) under Grant No 16302117 and partly by the Area of Excellence under the Grant No AoE/P-404/18-3. Both grants were issued by the Research Grants Council of Hong Kong S.A.R. This manuscript has been authored by Fermi Research Alliance, LLC, under Contract No. DE-AC02-07CH11359 with the U.S. Department of Energy, Office of Science, Office of High Energy Physics.

\newpage

\appendix
 \section{Event Response to $\Gamma_h$ Classifiers}
 \label{app:A}

\begin{figure}[ht]
	\centering
	\includegraphics[scale=.18]{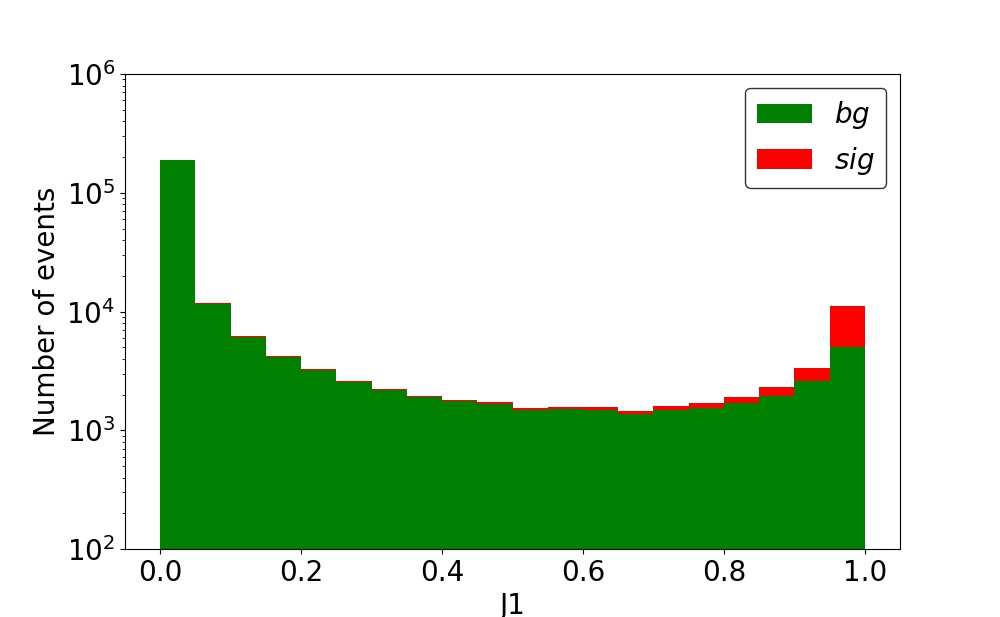} 
		\includegraphics[scale=.18]{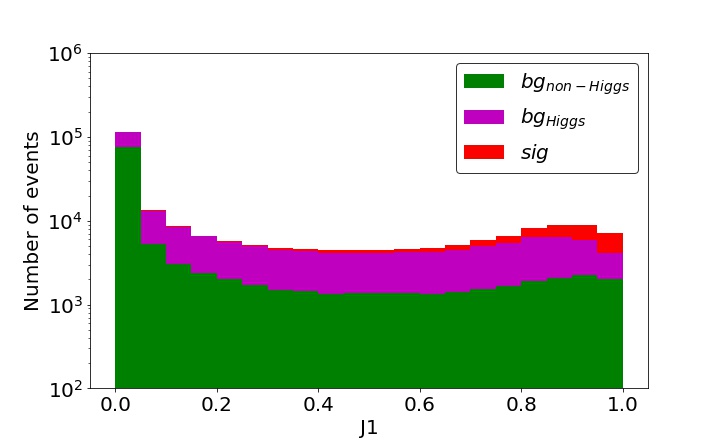} 	
		\includegraphics[scale=.18]{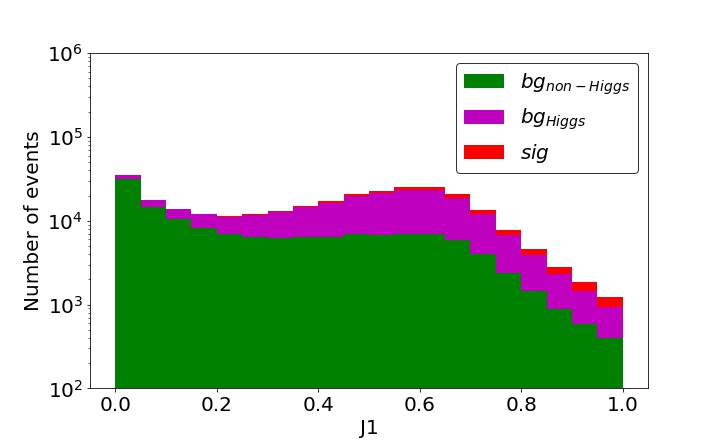}
		\includegraphics[scale=.18]{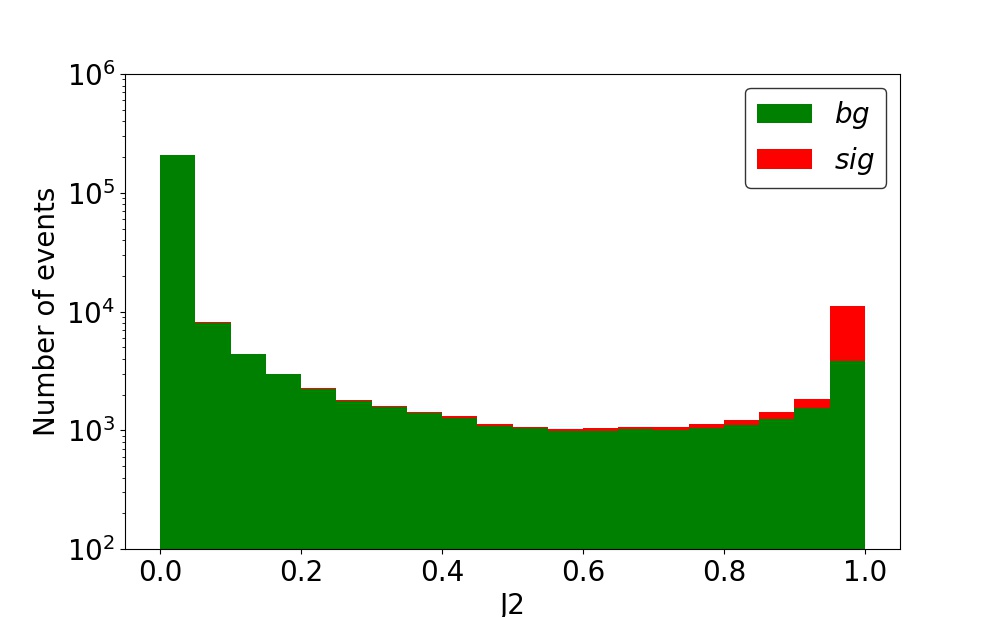}
				\includegraphics[scale=.18]{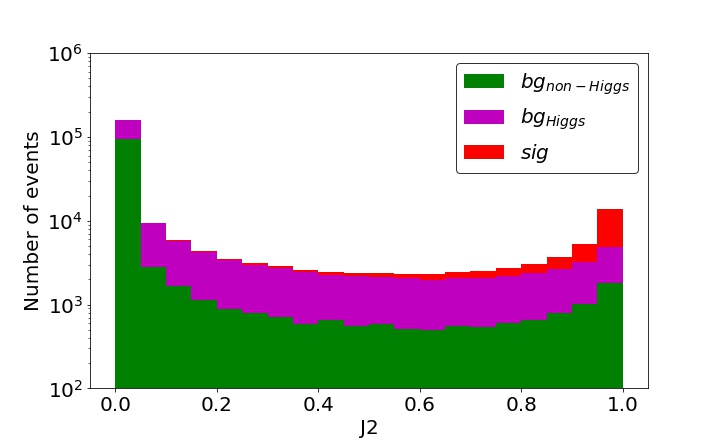}
				\includegraphics[scale=.18]{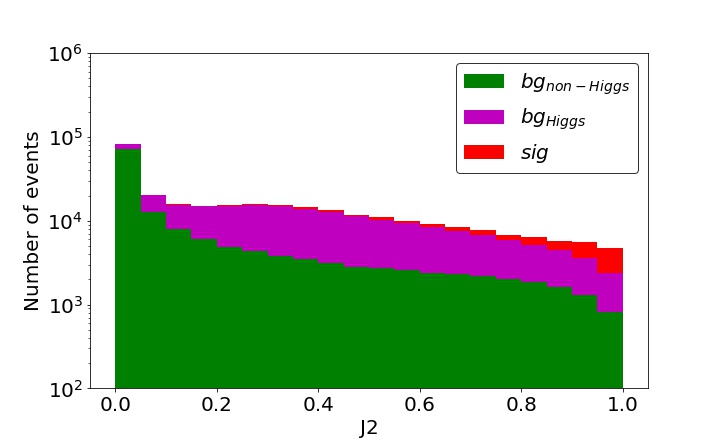}
			\includegraphics[scale=.18]{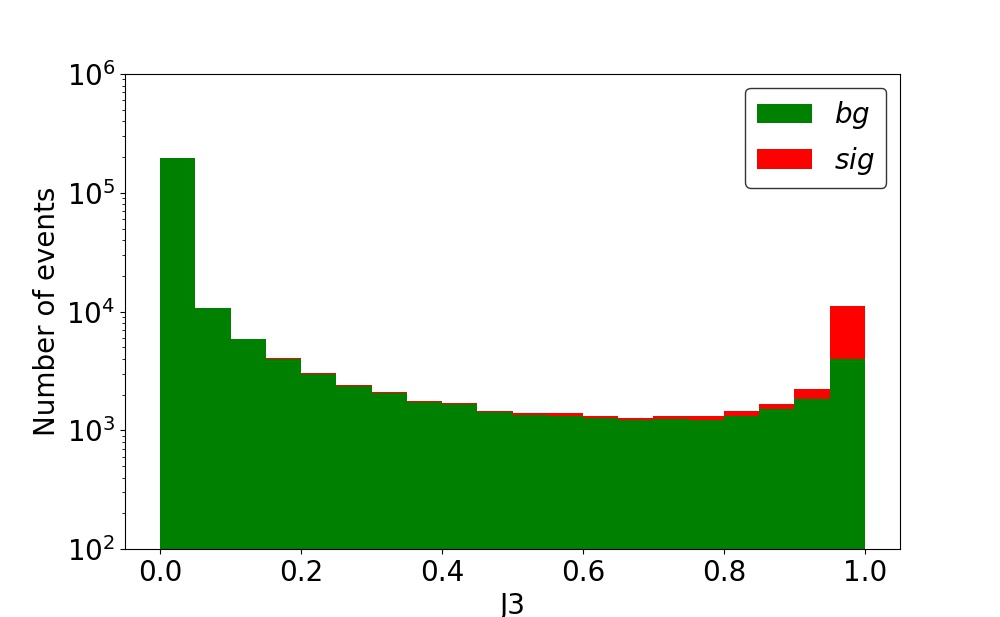} 
				\includegraphics[scale=.18]{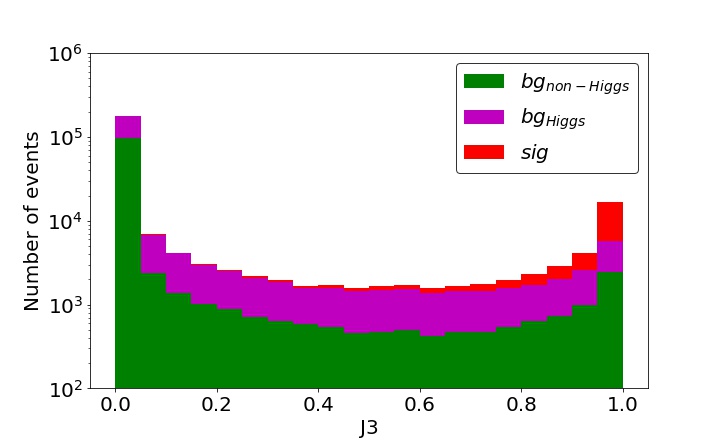} 
				\includegraphics[scale=.18]{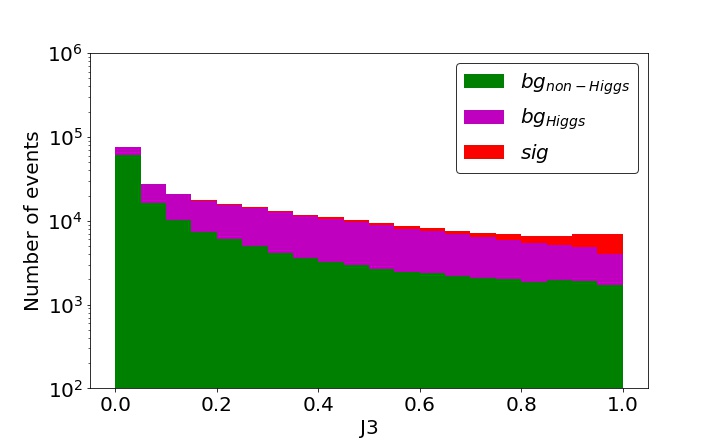} 
	\includegraphics[scale=.18]{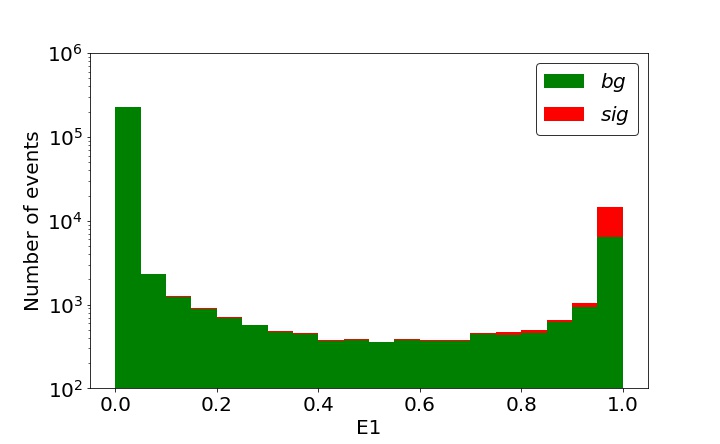}
		\includegraphics[scale=.18]{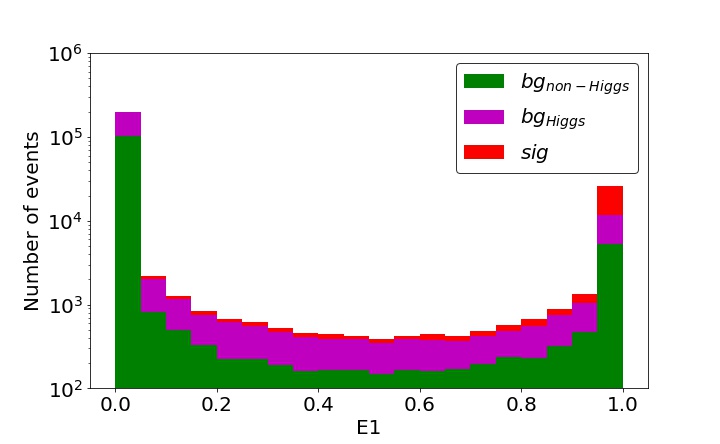}
		\includegraphics[scale=.18]{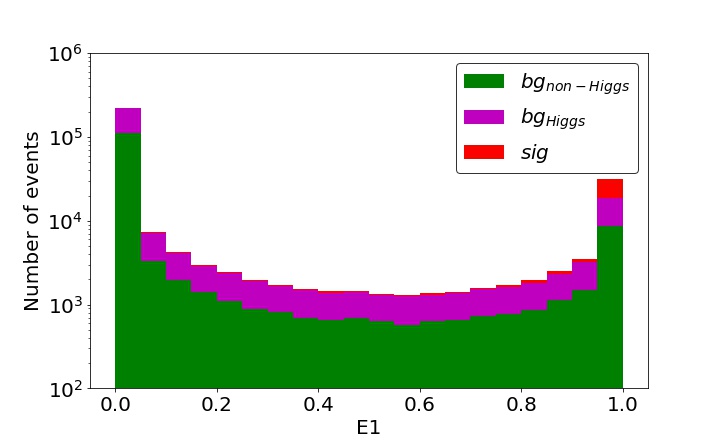}
	\includegraphics[scale=.18]{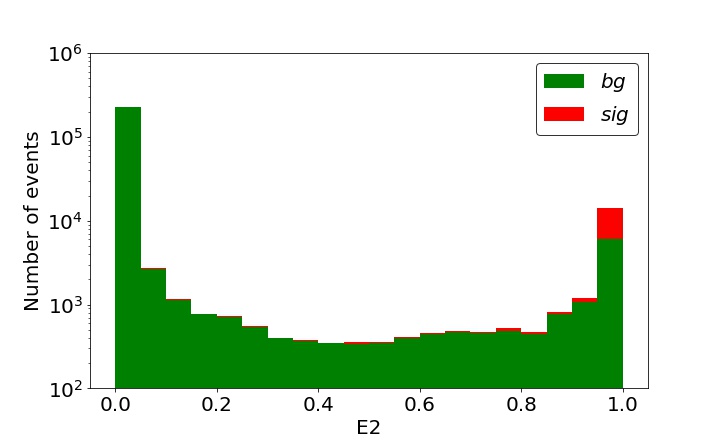}	
		\includegraphics[scale=.18]{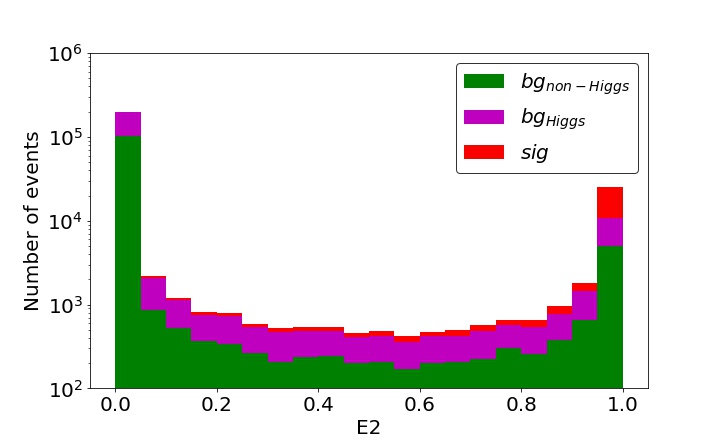}	
		\includegraphics[scale=.18]{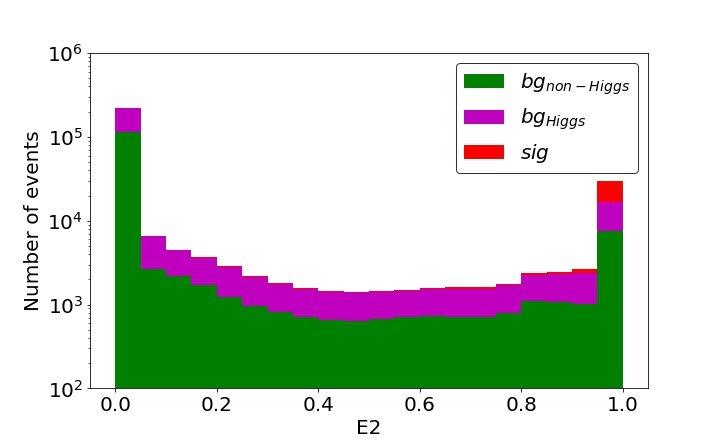}	
			\caption{Responses of the $Z_\nu h_{W_{lq}}$ (left), $Z_\nu h_{W_{qq}}$ (middle) and  $\nu\nu h_h$ (right) and their respective background events, to the binary classifiers.}
	\label{fig:eventvsML}
\end{figure}


\bibliographystyle{hieeetr}
\bibliography{reference}
\end{document}